\documentclass[a4paper,12pt]{article}
\usepackage{amsfonts,amsthm,amsmath,amssymb,graphicx}
\begin{document}

\newtheorem{definition}{Definition}[section]
\newcommand{\be}{\begin{equation}}
\newcommand{\ee}{\end{equation}}
\newcommand{\bea}{\begin{eqnarray}}
\newcommand{\eea}{\end{eqnarray}}
\newcommand{\LE}{\left[}
\newcommand{\R}{\right]}
\newcommand{\nn}{\nonumber}
\newcommand{\Tr}{\text{Tr}}
\newcommand{\N}{\mathcal{N}}
\newcommand{\G}{\Gamma}
\newcommand{\vf}{\varphi}
\newcommand{\LL}{\mathcal{L}}
\newcommand{\Op}{\mathcal{O}}
\newcommand{\HH}{\mathcal{H}}
\newcommand{\arctanh}{\text{arctanh}}
\newcommand{\up}{\uparrow}
\newcommand{\down}{\downarrow}
\newcommand{\ket}[1]{\left| #1 \right>}
\newcommand{\bra}[1]{\left< #1 \right|}
\newcommand{\ketbra}[1]{\left|#1\right>\left<#1\right|}
\newcommand{\rd}{\partial}
\newcommand{\de}{\partial}
\newcommand{\ba}{\begin{eqnarray}}
\newcommand{\ea}{\end{eqnarray}}
\newcommand{\db}{\bar{\partial}}
\newcommand{\we}{\wedge}
\newcommand{\ca}{\mathcal}
\newcommand{\lr}{\leftrightarrow}
\newcommand{\f}{\frac}
\newcommand{\s}{\sqrt}
\newcommand{\vp}{\varphi}
\newcommand{\hvp}{\hat{\varphi}}
\newcommand{\tvp}{\tilde{\varphi}}
\newcommand{\tp}{\tilde{\phi}}
\newcommand{\ti}{\tilde}
\newcommand{\ap}{\alpha}
\newcommand{\pr}{\propto}
\newcommand{\mb}{\mathbf}
\newcommand{\ddd}{\cdot\cdot\cdot}
\newcommand{\no}{\nonumber \\}
\newcommand{\la}{\langle}
\newcommand{\lb}{\rangle}
\newcommand{\ep}{\epsilon}
 \def\we{\wedge}
 \def\lr{\leftrightarrow}
 \def\f {\frac}
 \def\ti{\tilde}
 \def\ap{\alpha}
 \def\pr{\propto}
 \def\mb{\mathbf}
 \def\ddd{\cdot\cdot\cdot}
 \def\no{\nonumber \\}
 \def\la{\langle}
 \def\lb{\rangle}
 \def\ep{\epsilon}

\begin{titlepage}
\thispagestyle{empty}

\begin{flushright}
YITP-14-76\\
IPMU14-0311\\
\end{flushright}

\vspace{.4cm}
\begin{center}
\noindent{\Large \textbf{Quantum Entanglement of Localized Excited States at Finite Temperature}}\\
\vspace{2cm}

Pawe{\l} Caputa$^{a,b}$, Joan Sim\'on$^{c}$, Andrius \v{S}tikonas$^{c}$ and Tadashi Takayanagi$^{a,d}$
\vspace{1cm}

{\it
 $^{a}$Yukawa Institute for Theoretical Physics (YITP),\\
Kyoto University, Kyoto 606-8502, Japan\\
$^{b}$Nordita, KTH Royal Institute of Technology and Stockholm University,\\
Roslagstullsbacken 23, SE-106 91 Stockholm, Sweden\\
 $^{c}$ School of Mathematics and Maxwell Institute for Mathematical Sciences,\\
University of Edinburgh, King's Buildings,
Edinburgh EH9 3FD, UK\\
$^{d}$Kavli Institute for the Physics and Mathematics of the Universe (Kavli IPMU),\\
University of Tokyo, Kashiwa, Chiba 277-8582, Japan\\
}

\vskip 2em
\end{center}

\vspace{.5cm}
\begin{abstract}
In this work we study the time evolutions of (Renyi) entanglement entropy of locally excited states in two dimensional conformal field theories (CFTs) at finite temperature. We consider excited states created by acting with local operators on thermal states and give both field theoretic and holographic calculations. In free field CFTs, we find that the growth of Renyi entanglement entropy at finite temperature is reduced compared to the zero temperature result by a small quantity proportional to the width of the localized excitations. On the other hand, in finite temperature CFTs with classical gravity duals, we find that the entanglement entropy approaches a characteristic value at late time. This behaviour does not occur at zero temperature. We also study the mutual information between the two CFTs in the thermofield double (TFD) formulation and give physical interpretations of our results.

\end{abstract}

\end{titlepage}

\tableofcontents
%%%%%%%%%%%%%%%%%%%%%%%%%%
%%%%%%%%%%%%%%%%%%%%%%%%%%
\section{Introduction}
%%%%%%%%%%%%%%%%%%%%%%%%%%
%%%%%%%%%%%%%%%%%%%%%%%%%%
The AdS/CFT correspondence \cite{Maldacena} relates properties of quantum states in conformal field theories (CFTs) to those of geometries in their holographic spacetimes. The analysis of entanglement entropy in CFTs \cite{HLW,CC} and the holographic calculation of entanglement entropy \cite{RT} have shown that it is a useful quantity for connecting a spacetime geometry to its dual CFT data.

To probe the dynamical aspects of this correspondence by using entanglement entropy, it will be helpful to excite ground states in CFTs so that we can study propagations of excitations in the spacetime. So far, the most studied excited states are the so called quantum quenches, which are generated by a sudden change of Hamiltonian. We can choose, for example, this change to be translationally invariant and such quenches are called global \cite{cag}. The holographic analysis of global quenches has been discussed intensively (see e.g.\cite{QQ}).

An interesting construction of a gravity dual of global quench from an eternal AdS black hole has been found in \cite{HaMa}. In the extended Penrose diagram of eternal AdS black hole, there are two asymptotic AdS boundaries dual to two copies of CFTs in the thermofield formulation of finite temperature CFT. The gravity dual of quantum quench is obtained by removing one of the two boundaries and picking a half of eternal AdS black hole,  with an additional boundary put inside the horizon (see \cite{Caputa:2013eka} for generalization with chemical potential).
In the eternal AdS black hole dual to the thermofield formulation, it is also interesting to consider a time-dependent process when we excite one of the two CFTs. A related problem has been studied in the intriguing papers \cite{ShSt,Roberts:2014isa}. The consistency between the causality
of entanglement entropy and the bulk spacetime structure with such a non-trivial topology has been generally worked out in \cite{Headrick:2014cta}.

To probe local structures of spacetime, it will also be useful to consider locally excited states. A basic class of such locally excited states can be created by acting with local (primary field) operators $\mathcal{O}(x)$ on the CFT vacuum $|0\lb$.\footnote{Note that this class of excited states should be distinguished from the excited states which are given by primary states with definite eigenvalues of dilatational operator $(L_0,\bar{L}_0)$. See the appendix
\ref{prs} in the present paper for more details and the analysis of entanglement entropy.} The computations of (both von-Neumann and Renyi) entanglement entropies for such locally excited states have been formulated in \cite{Nozaki:2014hna} (see also \cite{UAM,Palmai:2014jqa} for a closely related calculations). Calculations of entanglement entropy for free scalar fields have been done in
\cite{Nozaki:2014hna,Nozaki:2014uaa} (see also \cite{Shiba} for another interesting approach). Results for rational conformal field theories in two dimensions have been obtained in \cite{He:2014mwa}. It was shown there that the increased amount of entanglement entropy is given by the log of the quantum dimension of the local operator. Moreover, computations of entanglement entropy for locally excited states in large N CFTs have been done in \cite{Caputa:2014vaa}.

In the paper \cite{NNT}, it was argued that a gravity dual for a locally excited state produced by a local operator with a large conformal dimension can be approximated by a falling massive particle in the AdS space (see also \cite{localq} for other approaches). The holographic entanglement entropy has been computed analytically in the AdS$_3/$CFT$_2$ set-up and the result is similar to the local quenches at zero temperature \cite{cal}. Recently, this holographic result has been reproduced from purely conformal field theoretic calculations using the large central charge limit by the authors in \cite{tom}. A generalization of this set-up to locally excited states in a finite temperature CFT would involve  a falling massive particle into an eternal AdS black hole. Therefore we expect that the analysis of entanglement entropy in such a set-up will be closely related to the physics of free falling objects in a black hole background.

Motivated by this, we would like to study the time-evolution of (Renyi) entanglement entropy of locally excited states in two dimensional CFTs at finite temperature. In particular, we study the Renyi entanglement entropy in both free field CFTs and large $N$ CFTs (or equally large central charge $c$ CFTs) by using the field theoretic replica method. In addition, we support our CFT results with the computation of the holographic entanglement entropy in a gravity dual consisting on the back-reacted geometry of a massive falling particle in a BTZ black hole. We also formulate the replica method calculation of Renyi entanglement entropy for the thermofield double description and compute the mutual information between two CFTs in the thermofield double (TFD).

This paper is organized as follows: In section two, we give a brief review of the
thermofield description of finite temperature CFTs and its holographic interpretation.  In section three, we compute the time-evolution of the Renyi entanglement entropy of locally excited states in two dimensional CFTs at finite temperature.
We also discuss our holographic results there. In section four, we formulate the replica method for the thermofield description, calculate the mutual information between two CFTs in the thermofield double and provide some intuitive explanation for our results. In section five, we summarize our conclusions. In appendix A, we study a class of excited states given by primary states in two dimensional CFTs and analyze their entanglement entropy. In appendix B, we present the exact expressions for the cross-ratios used in our CFT calculations.

When finishing this work, we became aware that an independent derivation of our results in appendix A was obtained in \cite{tom} as a part of their results.

%%%%%%%%%%%%%%%%%%%%%%%%%%%%%%%%%%%%%%%%%%%%%%%%%%%%
%%%%%%%%%%%%%%%%%%%%%%%%%%%%%%%%%%%%%%%%%%%%%%%%%%%%
\section{The thermofield double and black holes}
\label{sec:bulk}
%%%%%%%%%%%%%%%%%%%%%%%%%%%%%%%%%%%%%%%%%%%%%%%%%%%%
%%%%%%%%%%%%%%%%%%%%%%%%%%%%%%%%%%%%%%%%%%%%%%%%%%%%

Consider two non-interacting CFTs, say CFT$_L$ and CFT$_R$, in two dimensions with isomorphic Hilbert spaces $\mathcal H_{L,R}$. A particular {\it entangled state} in the total Hilbert space ${\cal H} = {\cal H}_L \otimes {\cal H}_R$ is the thermofield double state
\begin{equation}
\ket{\Psi_\beta}=\frac{1}{\sqrt{Z(\beta)}}\sum_n e^{-\frac{\beta}{2} E_n}\ket{n}_L\ket{n}_R
\label{eq:entangled}
\end{equation}
where $Z(\beta)=\sum_n e^{-\beta E_n}$ is the standard partition function in {\it one} of the Hilbert spaces. $\ket{n}_L$ is an eigenstate of the hamiltonian $H_L$ acting on $\mathcal{H}_L$ with eigenvalue $E_n$ (and similarly for $\ket{n}_R$). Furthermore, $\ket{n}_L$ is the CPT conjugate of the state $\ket{n}_R$ and to simplify notation we write $\ket{n}_L\otimes \ket{n}_R$ as $\ket{n}_L\ket{n}_R$.

By construction, the reduced density matrix of \eqref{eq:entangled} on either Hilbert space equals to a thermal state. For example, tracing over $\mathcal H_{L}$ gives rise to
\begin{equation}
  \rho_{R}(\beta) = \text{tr}_{\mathcal H_{L}} \left(\ket{\Psi_\beta}\bra{\Psi_\beta}\right) = \frac{1}{Z(\beta)}\sum_{n\in\mathcal H_{R}} e^{-\beta E_n} \ket{n}_R\bra{n}_R\,,
\end{equation}
the thermal state in $\mathcal H_{R}$. Thus, any correlation functions of observables $\mathcal{O}_R$ acting on $\mathcal H_{R}$ will equal thermal correlation functions
\begin{equation}
  \bra{\Psi_\beta} \mathcal{O}_R(x_1,t_1) \dots \mathcal{O}_R(x_n,t_n) \ket{\Psi_\beta} = \text{tr}_{\mathcal H_{R}} \left(\rho_{R}(\beta)  \mathcal{O}_R (x_1,t_1)\dots \mathcal{O}_R(x_n,t_n)\right)\,.
\label{eq:thermal-corr}
\end{equation}

Even in the absence of interactions, quantum entanglement is responsible for the existence of non-trivial correlations between ${\cal H}_L$ and ${\cal H}_R$. These correlations are encoded in {\it two-sided} correlation functions involving operators $\mathcal{O}_{L,R}$ acting on each Hilbert space $\mathcal H_{L,R}$, respectively,
\begin{equation}
  \bra{\Psi_\beta} \mathcal{O}_L(x_1,t) \dots \mathcal{O}_R(x^\prime_n,t^\prime_n) \ket{\Psi_\beta}\,.
\label{eq:two-sided}
\end{equation}
Remarkably, these two-sided correlators can be computed by analytical continuation
\begin{equation}
  \bra{\Psi_\beta} \mathcal{O}_L(x_1,-t) \dots \mathcal{O}_R(x^\prime_n,t^\prime_n) \ket{\Psi_\beta} = \text{tr}_{\mathcal H_{R}} \left(\rho_{R}(\beta)  \mathcal{O}_R(x_1,t-i\beta/2) \dots \mathcal{O}_R(x^\prime_n,t^\prime_n)\right)\,.
\label{eq:anal-cont}
\end{equation}
This observation will play an important role in our CFT entanglement calculations in section~\ref{MutualInformation}.

%%%%%%%%%%%%%%%%%%%%%%
\subsection{Gravity dual description}
%%%%%%%%%%%%%%%%%%%%%%

\begin{figure}[h!]
\centering
\includegraphics[width=6cm]{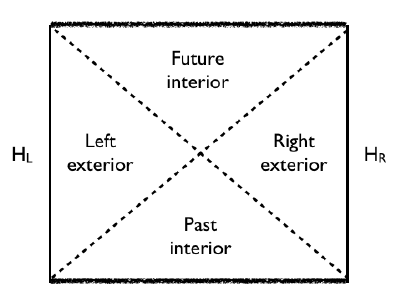}
\caption{Penrose diagram for an eternal AdS black hole.
\label{fig:eternal}}
\end{figure}

Whenever the 2d CFTs in the previous discussion have a holographic dual, the AdS/CFT correspondence asserts the existence of
a gravity dual realization of the thermofield double state. Maldacena proposed the entangled state \eqref{eq:entangled} to be dual to the eternal AdS black hole \cite{MaBH}. For 2d CFTs, this would correspond to the BTZ black hole \cite{BTZ}. Its Penrose diagram is shown in figure~\ref{fig:eternal}.

The existence of two conformal boundaries matches the presence of two CFTs in our field theory discussion. Tracing over $\mathcal{H}_L$ is equivalent to tracing over the region of spacetime causally connected to it. This is why an observer at infinity, measuring in $\mathcal{H}_R$, perceives her event horizon as a thermal atmosphere. This is in manifest agreement with why observables measured by such observers are thermal.

There are two bulk interpretations of the thermofield double state \eqref{eq:entangled}. The most canonical one describes a {\it single} black hole in thermal equilibrium \cite{israel}. Hamiltonian evolution is generated by the boost
\begin{equation}
  H_\text{tf} = \mathbb{I}_L\otimes H_R - H_L\otimes \mathbb{I}_R\,.
\end{equation}
To simplify our notation, we will refer to it as $H_R-H_L$. Its action can be understood as propagating time upwards in $\mathcal{H}_R$ and downwards in $\mathcal{H}_L$. This is again in agreement with \eqref{eq:anal-cont}. Notice the state \eqref{eq:entangled} is an eigenstate of this hamiltonian with vanishing eigenvalue. Thus, the state has no time evolution because it is boost invariant.

An alternative interpretation of the eternal AdS black hole is as an approximate description of the state at $t=0$ of {\it two} AdS black holes. In this case, hamiltonian evolution is generated by
\begin{equation}
  H = \mathbb{I}_L\otimes H_R + H_L\otimes \mathbb{I}_R \equiv H_R+H_L\,.
\end{equation}
Notice this corresponds to propagating time upwards in both boundaries. Since this action is no longer an isometry of the geometry, the thermofield double state  \eqref{eq:entangled} will have a non-trivial time evolution. In fact, it is no longer an eigenvector of $H$, but instead
\begin{equation}
  \ket{\Psi_\beta(t)}=\frac{1}{\sqrt{Z(\beta)}}\sum_n e^{-\frac{\beta}{2} E_n}\,e^{-2iE_n t}\,\ket{n}_L\ket{n}_R
\end{equation}
Despite the manifest time dependence of the state, the corresponding reduced density matrices ($\rho_{L}$ or $\rho_{R}$)  remain time-independent thermal density matrices.

If the quantum state would be a product state, then the two black holes would be described by disconnected spaces, in agreement with the lack of correlations between $CFT_L$ and $CFT_R$. The existence of 2-sided correlations in the CFT, due to quantum entanglement of the thermofield double state \eqref{eq:entangled}, can be generically understood in terms of causal connection between the two boundaries in the future interior of the black hole. More recently, it was suggested in \cite{juanlenny} that even though the two black holes belong to non-interacting universes in the second bulk interpretation of the thermofield double state, the manifestation of quantum entanglement is through an Einstein-Rosen bridge connecting both geometries : one identifies their bifurcate horizons and fills in the space in the interior of both black holes. This reasoning lead to the EPR=ER conjecture \cite{juanlenny}.

One of the motivations of this work is to develop the CFT machinery required to understand the physics of restoration to thermal equilibrium, after perturbing the thermal equilibrium state with a local perturbation, using entanglement measures. This has intrinsic value in different branches of physics. For example, from the perspective of holography and black hole physics, one of our motivations is to eventually carry first principle CFT calculations that will be able to probe the physics of {\it scrambling} \cite{lenny-1}. The gravity dual calculation has been recently done in \cite{ShSt,Roberts:2014isa}, where the authors compute the mutual information and other correlations in shock wave geometries.

%%%%%%%%%%%%%%%%%%%%%%%%%%%%%%%%%%%%%%%%%%%%
%%%%%%%%%%%%%%%%%%%%%%%%%%%%%%%%%%%%%%%%%%%%
\section{(Renyi) entanglement entropy for locally excited states at finite temperature}
\label{sec:thermal}
%%%%%%%%%%%%%%%%%%%%%%%%%%%%%%%%%%%%%%%%%%%%
%%%%%%%%%%%%%%%%%%%%%%%%%%%%%%%%%%%%%%%%%%%%

\begin{figure}[h!]
\centering
\includegraphics[width=8cm]{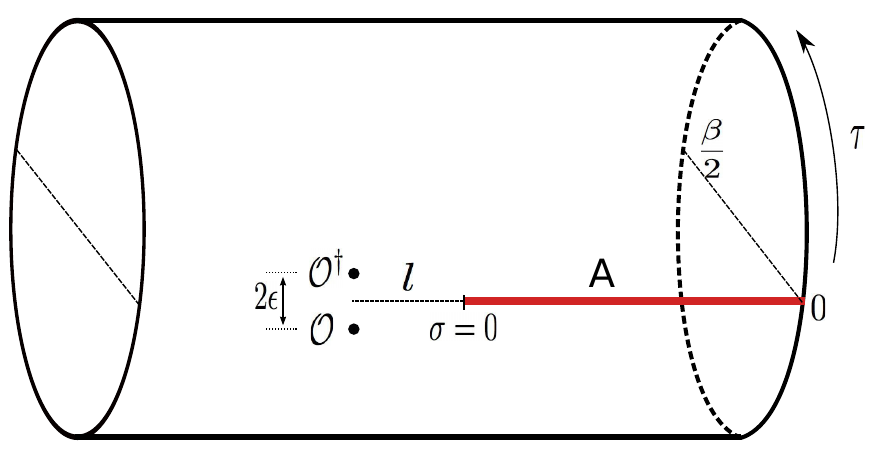}
\caption{The reduced density matrix for our excited states is described by a path-integral on a cylinder with complex coordinates $x=\sigma+i\tau$. The two operators inserted at distance $x=-l$ from the cut $A$ are separated by $\Delta x=2i\epsilon$.  $\Tr \rho^n_A$ is computed as a partition function on the n-copies of these cylinders glued along $A$.
\label{fig:Cyl1}}
\end{figure}

Consider a 2d CFT in a thermal state and locally perturb it by inserting an operator $\mathcal{O}$ at $t=0$, $x=-l$. Hamiltonian evolution gives rise to the time dependent density matrix
\begin{equation}
\begin{aligned}
\rho(t)&=\mathcal{N}e^{-iHt}e^{-\epsilon H}\mathcal{O}(-l)e^{-\beta H+2\epsilon H}\mathcal{O}^{\dagger}(-l) e^{-\epsilon H}e^{iHt} \\
&\equiv\mathcal{N}\mathcal{O}(x_2,\bar{x}_2)\,e^{-\beta H}\,\mathcal{O}^{\dagger}(x_1,\bar{x}_1)
\end{aligned}
\label{DMDS}
\end{equation}
where $\mathcal{N}$ is a normalization constant to ensure $\Tr(\rho)=1$. We are following the same notation as in \cite{Nozaki:2014hna,He:2014mwa,Caputa:2014vaa}. In particular, the $\epsilon$ factor was introduced to regulate the distance between the insertion points of the local operators (Fig.\ref{fig:Cyl1}). More precisely, these points are
\bea
x_1&=&t-l+i\epsilon\qquad \bar{x}_1=-l-t-i\epsilon\\
x_2&=&t-l-i\epsilon\qquad \bar{x}_2=-l-t+i\epsilon\,.\label{Points}
\eea
Thus, the distance $x_1-x_2$ equals $2i\epsilon$.

We are interested in measuring the amount of entanglement in some region $A$. For simplicity, we will focus on the semi-infinite interval $A=[0,\infty]$. By tracing out the complement of region $A$, we obtain the reduced density matrix $\rho_A(t)$. Then, using the replica method and following \cite{Nozaki:2014hna}, we define the growth of the $n$-th Renyi entanglement entropy in a thermal state excited by a local operator as
\be
\Delta S^{(n)}_{A}=\frac{1}{1-n}\log\left(\frac{\Tr(\rho^n_A)}{\Tr(\rho^{(0)}_A)^n} \right)=\frac{1}{1-n}\log\left[\frac{\langle \mathcal{O}(x_1,\bar{x}_1)\mathcal{O}^{\dagger}(x_2,\bar{x}_2)...\mathcal{O}^{\dagger}(x_{2n},\bar{x}_{2n})\rangle_{C_n}}{\left(\langle \mathcal{O}(x_1,\bar{x}_1)\mathcal{O}^{\dagger}(x_2,\bar{x}_2)\rangle_{C_1}\right)^n}\right]\label{Renyin}
\ee
The $2n$-point function is computed on an $n$-sheeted cylinder $C_n$, each with a cut corresponding to $A$, glued cyclically along the semi-infinite intervals. A pair of operators is inserted on each cylinder, to describe the local excitation, and separated by $2i\epsilon$. The insertion points on the $n$-th cylinder are
\be
x_{2n-1}=x_1+i(n-1)\beta,\qquad x_{2n}=x_2+i(n-1)\beta
\ee
$\rho^{(0)}_A$ is the density matrix {\it without} the insertion of local operators. Thus, by construction, $\Delta S^{(n)}_{A}$ measures the variation in the $n$-th Renyi entanglement entropy in $A$ due to the local excitations $\mathcal{O}$.

From now on, we focus on the second Renyi entanglement entropy (i.e. $n=2$) and on excitations by primary operators just for the simplicity of computations. $\Delta S^{(2)}_A$ depends both on the 2-pt function on the cylinder,
\be
\langle \mathcal{O}(x_1,\bar{x}_1)\mathcal{O}^{\dagger}(x_2,\bar{x}_2)\rangle_{C_1}=\left|\frac{\beta}{\pi}\sinh\left(\frac{\pi x_{12}}{\beta}\right)\right|^{-4\Delta_O}
\label{eq:2ptT}
\ee
and a 4-pt function on $C_2$. The latter can be computed if we find a conformal map from $C_2$ to the complex plane $\mathbb{C}$, where such 4-pt function is given by
\be
\langle \mathcal{O}^{\dagger}(z_1,\bar{z}_1)\mathcal{O}(z_2,\bar{z}_2)\mathcal{O}^{\dagger}(z_3,\bar{z}_3)\mathcal{O}(z_4,\bar{z}_4)\rangle_{\mathbb{C}}=|z_{13}z_{24}|^{-4\Delta_O}G(z,\bar{z})\,,
\label{4pGen}
\ee
where $z_{ij} = z_i-z_j$ and $G(z,\bar{z})$ is a theory dependent function of the conformal ratios $z=(z_{12}z_{34})/(z_{13}z_{24})$.

The crucial step in our calculation is to realise that the two-sheeted cylinder $C_2$ with coordinates $x=\sigma+i\tau$ and semi-infinite cuts at $\tau=0$ can be mapped to the complex plane by
\be
z(x)=\sqrt{e^{\frac{2\pi x}{\beta}}-1}
\label{MapSI}
\ee
This map can be derived as a composition of conformal maps. Indeed, first, map each cylinder with a semi-infinite cut to the plane with an interval cut from $[1,\infty]$ using the exponential map $w(x)=\exp(2\pi x/\beta)$. Second, use the uniformization map $z^2(w)=w-1$\footnote{There exists a similar map $z^2(x)=(w(x)-1)/(w(x)-w(L))$ for a cylinder with finite interval cut $[0,L]$ to the complex plane $\mathbb{C}$. For clarity of the presentation we only consider $A\in [0,\infty]$.}.

Using the map \eqref{MapSI}, we learn that the insertion points of the operators in the second cylinder are mapped to the plane points
$z_3=z(x_1+i\beta)=-z_1$ and $z_4=z(x_2+i\beta)=-z_2$. This implies that the cross-ratio determining $G(z,\bar{z})$ reduces to
\be
z_A\equiv \frac{z_{12}z_{34}}{z_{13}z_{24}}=\frac{1}{2}\left(1-\frac{z^2_1+z^2_2}{2z_1z_2}\right)
\ee
and similarly for $\bar{z}_A$, where the subindex $A$ refers to the dependence on the semi-infinite cut characterising the entanglement region $A$.

Using the transformation law of correlators of primary operators under conformal maps, it is straightforward to show that
\be
\frac{\langle \mathcal{O}(x_1,\bar{x}_1)\mathcal{O}^{\dagger}(x_2,\bar{x}_2)\mathcal{O}(x_3,\bar{x}_3)\mathcal{O}^{\dagger}(x_4,\bar{x}_4)\rangle_{C_2}}{\left(\langle \mathcal{O}(x_1,\bar{x}_1)\mathcal{O}^{\dagger}(x_2,\bar{x}_2)\rangle_{C_1}\right)^2}=\left|z_A(1-z_A)\right|^{4\Delta_O}G(z_A,\bar{z}_A)\,.
\ee
This is a general formula for a 2d CFT that only relies on the conformal map \eqref{MapSI}. It leads to a concise expression for the growth of the Renyi entanglement entropy
\be
\Delta S^{(2)}_A=-\log\left(\left|z_A(1-z_A)\right|^{4\Delta_O}G(z_A,\bar{z}_A)\right)\,.
\label{DSmain}
\ee
Let us stress that the dependence of $\Delta S^{(2)}_A$ on time, the inverse temperature $\beta$ and the regulator $\epsilon$ is entirely through the cross-ratios $z_A$ and $\bar{z}_A$.

The examination of these cross-ratios for the map \eqref{MapSI} is explained in appendix~\ref{App:Details}. The main observation is that if we remove the cut-off, i.e. $\epsilon\to 0$, we recover the same behaviour reported for $T=0$ in \cite{Nozaki:2014hna,Nozaki:2014uaa,He:2014mwa,Caputa:2014vaa}. This is because the cross-ratios behave as $(z_A,\bar{z}_A)\to(0,0)$ for early times $t<l$, whereas we get $(z_A,\bar{z}_A)\to(1,0)$ for late times $t>l$. Thus, at late time, the growth of the Renyi entanglement entropy approaches a constant value\footnote{For a finite interval entanglement region of size $L$, the constant value is obtained for $l<t<l+L$.}. The latter characterizes the entanglement of local operators.

This conclusion may not be surprising since the 2-pt function at finite temperature \eqref{eq:2ptT} is not sensitive to the temperature in the deep UV. Since our primary operators are inserted $2i\epsilon$ away, the $\epsilon\to 0$ limit explores such UV and the answer is expected to be temperature independent. Thus, to take into account finite temperature effects, we must keep the regulator $\epsilon$ finite. We elaborate on this observation below in a few examples. We will provide a physical interpretation for these finite temperature effects in coming sections.

%%%%%%%%%%%%%%%%%%%%%%%%%%%%%%%%
\subsection{Finite $\epsilon$ examples}
%%%%%%%%%%%%%%%%%%%%%%%%%%%%%%%%

In this section we consider three different situations : free bosons, the Ising model and the large $c$ limit of 2d CFTs. In all our calculations of  $\Delta S^{(2)}_A$, we will keep $\epsilon$ small, but finite, keeping only terms up to order $O(\epsilon)$.

\paragraph{Free bosons :} Our first example considers the CFT of a single free boson $\phi(z,\bar{z})$ in two dimensions. We study this CFT in the thermal state \eqref{DMDS} with a local primary operator
\be
\mathcal{O}(x,\bar{x})=\frac{1}{\sqrt{2}}\left(e^{\frac{i}{2}\phi}+e^{-\frac{i}{2}\phi}\right),\qquad \Delta_O=\bar{\Delta}_O=\frac{1}{8}\label{OpFS}
\ee
At zero temperature, the evolution of $\Delta S^{(n)}_A$ in a state locally excited by this operator was studied in
\cite{He:2014mwa}. It was found that $\Delta S^{(n)}_A\to \log 2$ at late time. Following \cite{cal}, the physical interpretation is as follows.
The insertion of the local operator \eqref{OpFS} is equivalent to the creation of an entangled EPR-like pair propagating with the speed of light away from the insertion points. Once one member of the pair enters the region $A$, it contributes to the entanglement between $A$ and its complement by a value of $\log 2$, a bit of entanglement. Let us see how this picture is modified at finite temperature.

For a free boson 2d CFT, the function $G(z,\bar{z})$ controlling 4-pt functions equals \cite{He:2014mwa}
\be
G(z,\bar{z})=\frac{1}{2\sqrt{|z(1-z)|}}\left(1+|z|+|1-z|\right)\,.
\label{GFS}
\ee
Inserting this into \eqref{DSmain} we obtain the growth of the second Renyi entanglement entropy for a free boson CFT with local EPR-like operator \eqref{OpFS}
\be
\Delta S^{(2)}_A=\log\left(\frac{2}{1+|z_A|+|1-z_A|}\right)
\ee
A careful analysis of the cross-ratios $z_A,\,\bar{z}_A$ \eqref{CRSI} for small, but finite $\epsilon$, reveals that at late time $t>l$, the limiting constant value for the growth of the Renyi entangelment entropy is decreased as follows
\be
\Delta S^{(2)}_A\simeq \log2-\frac{\pi\epsilon}{\beta}+O(\epsilon^2)
\ee
Physically, this can be understood by the influence of thermal fluctuations on the propagating EPR pair that is now represented by two wave packets of size $\epsilon$ propagating away from each other\footnote{We comment further on this wave packet interpretation in section~\ref{sec:density}.}.

\paragraph{Critical Ising model :} The second example is the critical Ising model, i.e. the $(4,3)$ minimal model also studied in \cite{He:2014mwa}. Consider the spin $\sigma$ primary operator as a local excitation
\be
\mathcal{O}(x,\bar{x})=\sigma(x,\bar{x}),\qquad \Delta_{\sigma}=\bar{\Delta}_{\sigma}=\frac{1}{16}
\ee
Notice the chiral dimension of $\sigma$ is half that of the free scalar operator \eqref{OpFS} ($\Delta_{\sigma}=\frac{1}{2}\Delta_O$).
Furthermore, a general 2n-point correlator in the critical Ising model on the plane is equal to the square root of the 2n-point function of the free boson operator \cite{DiFrancesco:1987ez}. As a result, the growth of the Renyi entanglement entropy is equal to $1/2$ of the free boson answer. For small $\epsilon$, this equals
\be
\Delta S^{(2)}_A\simeq \log\left(\sqrt{2}\right)-\frac{\pi\epsilon}{2\beta}+O(\epsilon^2)
\ee
This is the logarithm of the quantum dimension of $\sigma$ obtained in \cite{He:2014mwa}, but decreased by the finite temperature effect, which equals half of the amount in the free boson CFT calculation. This is still consistent with the picture of the propagating EPR-like pair since different operators can contribute different amounts to the entanglement entropy.

\paragraph{Large $c$ general 2d CFT:}  Consider a two dimensional CFT with a large central charge $c$
which has a gap as required by the gravity dual as in the calculation of \cite{gap}. We take a locally excited state and assume that the conformal dimension of local operator $\Delta_O$ is much larger than one but much smaller than $c$:
$1\ll \Delta_O \ll c$. As shown in \cite{Caputa:2014vaa}, at zero temperature the growth of the Renyi entanglement entropy diverges logarithmically with time with a universal coefficient proportional to the conformal dimension $\Delta_O$ of the local operator used as the excitation
\be
\Delta S^{(2)}_A\simeq 4\Delta_O\log\left(\frac{2t}{\epsilon}\right)\label{logt}
\ee
This CFT prediction \eqref{logt} was reproduced holographically in the large $\Delta_O$ limit corresponding to the geodesic bulk approximation to correlators \cite{Caputa:2014vaa}, keeping only disconnected Witten diagrams.

A gravity dual for heavy local operator insertion was proposed in \cite{NNT}, which is given by a massive particle falling from the AdS boundary to the interior horizon. The corresponding calculations of holographic entanglement entropy lead to the behavior $\Delta S_A\sim \f{c}{6}\log \f{t}{\ep}$. This holographic result has been fully reproduced from a large $c$ limit analysis in \cite{tom} for the operators such that $\Delta_O\sim O(c)$.
 This logarithmic time dependence is reminiscent of the one found in local quenches \cite{cal} after replacing $\Delta_O$ by the central charge $c$ of the CFT, which behaves like $\Delta S_A\sim \f{c}{3}\log \f{t}{\ep}$.

To study the effect of finite temperature, we use the approximation for the conformal block in the large $c$ limit \cite{Fateev:2011qa}
\be
G(z,\bar{z})\simeq |z|^{-4\Delta_O}\,.
\label{Fateev}
\ee
Since the latter is valid at all times, the growth of the Renyi entanglement entropy \eqref{DSmain} becomes
\be
\Delta S^{(2)}_A\simeq -4\Delta_O\log\left(|1-z_A|\right)
\label{LargeCDS}
\ee
We stress this is a {\it universal} 2d CFT result in the large $c$ limit. Cross-ratios  \eqref{CRSI} determine the behaviour of this Renyi entropy : $z_A,\,\bar{z}_A\to 0$ for $t<l$, leading to a vanishing growth in the Renyi entanglement entropy, whereas $z_A\to 1$, keeping $\bar{z}_A\to 0$
for $t>l$. We learn that, at late times and small $\epsilon$
\bea
  \Delta S^{(2)}_A&\simeq&4\Delta_O\log\left(\frac{\beta}{\pi\epsilon}\right)+O(\epsilon^2)\,.
\label{RELc}
\eea
Thus, the effect of finite temperature is equivalent to introducing a cut-off for the time $t_{\max}\simeq \beta/2\pi$. A CFT at finite temperature has a black hole geometry as a holographic dual \cite{Witten:1998zw}. It is then very suggestive to interpret this time cut-off derived in the CFT, as the time taking a massive particle to approach the black hole horizon. We will discuss this interpretation in subsection \ref{sec:backreaction} when presenting our gravity dual calculations.

%%%%%%%%%%%%%%%%%%%%%%%%%%%%%%%%%%%%%%%%%%%%%%%%%%%%
\subsection{Energy density}
\label{sec:density}
%%%%%%%%%%%%%%%%%%%%%%%%%%%%%%%%%%%%%%%%%%%%%%%%%%%%

As shown in \cite{NNT}, the picture of propagating EPR pairs is also consistent with the expectation value of the energy. Namely, the time dependent correlator of the energy has the form of two profiles propagating from the insertion point of the local operator. Let us then compute the expectation value of  $T_{tt}(x,\bar{x})=-(T(x)+\bar{T}(\bar{x}))$ in the excited state in 2d CFT at finite temperature
\be
\langle T_{tt}(x,\bar{x})\rangle_\mathcal{O}\equiv\frac{\langle \mathcal{O}^{\dagger}(x_1,\bar{x}_1)T_{tt}(x,\bar{x})\mathcal{O}(x_2,\bar{x}_2)\rangle_{C_1}}{\langle \mathcal{O}^{\dagger}(x_1,\bar{x}_1)\mathcal{O}(x_2,\bar{x}_2)\rangle_{C_1}}.
\label{eq:energy}
\ee
Using the map from the plane to the cylinder with circumference $\beta$, $z=\exp\left(\frac{2\pi x}{\beta}\right)$ and the appropriate transformation of the primary operators as well as the energy momentum tensor we get
\be
\langle T_{tt}(x,x)\rangle_\mathcal{O}=\frac{c\pi^2}{3\beta^2}-\frac{\frac{\pi^2\Delta_O}{\beta^2}\sinh^2\left(\frac{\pi x_{12}}{\beta}\right)}{\sinh^2\left(\frac{\pi (x-x_1)}{\beta}\right)\sinh^2\left(\frac{\pi (x-x_2)}{\beta}\right)}-\frac{\frac{\pi^2\Delta_O}{\beta^2}\sinh^2\left(\frac{\pi \bar{x}_{12}}{\beta}\right)}{\sinh^2\left(\frac{\pi (x-\bar{x}_1)}{\beta}\right)\sinh^2\left(\frac{\pi (x-\bar{x}_2)}{\beta}\right)}
\label{eq:genergy}
\ee
Once we plug the insertion points points \eqref{Points} the final result becomes
\begin{equation}
\begin{aligned}
\langle T_{tt}(x)\rangle_\mathcal{O} &=
   \frac{\frac{4 \pi ^2 \Delta_O}{\beta ^2}\sin ^2\left(\frac{2 \pi  \epsilon }{\beta
   }\right)}{\left(\cosh \left(\frac{2 \pi  (l-t+x)}{\beta
   }\right)-\cos \left(\frac{2 \pi  \epsilon }{\beta
   }\right)\right)^2}+\frac{\frac{4 \pi ^2 \Delta_O}{\beta ^2}\sin ^2\left(\frac{2 \pi
   \epsilon }{\beta }\right)}{\left(\cosh \left(\frac{2 \pi
   (l+t+x)}{\beta }\right)-\cos \left(\frac{2 \pi
   \epsilon }{\beta }\right)\right)^2}\\
&+\frac{\pi ^2 c}{3 \beta ^2}\,.
\end{aligned}
\label{eq:local}
\end{equation}

The expectation value of the energy is then described by two profiles propagating away from each other from the insertion of the operator $x=-l$. The width of the profiles is determined by the ratio of $\epsilon/\beta$ and in the limit $\epsilon\to 0$ the profiles become two delta functions as in the zero temperature case \cite{NNT}.

 A basic property of our excited state created by the primary operator $\mathcal{O}$ insertion can be seen from \eqref{eq:local} : the energy density is that of a thermal state $\left(\frac{\pi ^2 c}{3 \beta ^2}\right)$ plus two lumps of energy propagating in opposite directions. Thus, there is no natural sense in which this perturbation relaxes to thermal equilibrium.

%%%%%%%%%%%%%%%%%%%%%%%%%%%%%%%%%%%%%%%%%%%%%%%%%%%%
\subsection{Holographic results: falling particle in BTZ}
\label{sec:backreaction}
%%%%%%%%%%%%%%%%%%%%%%%%%%%%%%%%%%%%%%%%%%%%%%%%%%%%
In this section we consider the entanglement entropy $S_A$ for the subsystem $A$ defined by the semi-infinite line  $x>0$ in a two dimensional CFT with a gravity dual. Based on the AdS$_3/$CFT$_2$, we want to holographically compute the time evolution of the entanglement entropy after the local operator insertion at $x=0$, by simply setting $l=0$ in our previous set-up.

In the AdS/CFT correspondence, the perturbation due to a heavy operator with conformal dimension $\Delta(=2\Delta_O)$ can be approximated by a massive point particle with mass $m=\Delta/R$ starting its motion at the distance $z=\epsilon$  from the boundary in Poincar\'e AdS space with AdS radius $R$ and, as time progresses, falls into the AdS horizon \cite{NNT}. The particle back-reaction is initially localised around the particle and spreads out with time. The mass parameter $\mu$ characterising the back-reacted gravity solution is proportional to the mass $m$ of the falling particle, $\mu=8G_N R^2\,m$, In terms of the CFT dual data, one finds the relation $\mu=\f{24\Delta_O}{c}R^2$.

\paragraph{Geodesic approximation :} A natural generalisation of this set-up to finite temperature is to study a falling particle in the BTZ background \cite{BTZ}
\be
ds^2=\frac{R^2}{z^2}\left(-\left(1-Mz^2\right)dt^2+\frac{dz^2}{\left(1-Mz^2\right)}+dx^2\right)\,.
\label{BTZ}
\ee
The mass $M$ of the black hole is related to its Hawking's temperature by $\beta=T^{-1}=\frac{2\pi}{\sqrt{M}}$. Furthermore, since our CFT calculations involved a non-compact manifold,  we will take $x\in\mathbb{R}$. Thus,  we will be considering the BTZ string background, as in \cite{HaMa}. We parametrize the trajectory of a particle at $x=0$ in the gauge $(t,z)=(\tau,z(\tau))$. The action for such particle of mass $m$ in \eqref{BTZ} is then given by
\be
S_p=-mR\int \frac{d\tau}{z(\tau)}\sqrt{1-Mz(\tau)^2-\frac{\dot{z}(\tau)^2}{1-Mz(\tau)^2}}\,.
\ee
Its equations of motion yield the trajectory
\be
z(\tau)=\frac{\beta}{2\pi}\sqrt{1-\left(1-\left(\frac{2\pi\epsilon}{\beta}\right)^2\right)\left(1-\tanh^2\left(\frac{2\pi\tau}{\beta}\right)\right)}\,.
\ee
Notice we already used the boundary condition $z(0)=\epsilon$, where $\epsilon$ parametrises the size of the CFT perturbation, as in our previous subsection. The behaviour of this geodesic already indicates that the
natural time scale after which the particle is {\it close} to the horizon, $z\sim 1/\sqrt{M} = \beta/(2\pi)$, is $t\simeq \frac{\beta}{2\pi}$, already matching our CFT calculation in \eqref{RELc}.

%%%%%%%%%%%%%%%%%%
\paragraph{Back-reaction metric:} The back-reaction of this falling massive particle is found in complete analogy to \cite{Horowitz:1999gf,NNT}. Since the problem can be solved in global AdS${}_3$, all we have to do is to map the BTZ coordinates into global ones. We provide this map below using the
embedding $\mathbb{R}^{2,2}$ space :
\begin{equation}
\begin{aligned}
\sqrt{r^2+R^2}\sin\tilde{\tau}&=\frac{R}{\sqrt{M} z}\sqrt{1-Mz^2}\sinh \left(\sqrt{M}t\right)\,, \\
\sqrt{r^2+R^2}\cos\tilde{\tau}&=\frac{R\left(\cosh(\lambda)\cosh\left(\sqrt{M} x\right)-\sqrt{1-Mz^2}\sinh(\lambda)\cosh\left(\sqrt{M}t\right)\right)}{\sqrt{M}z}\,, \\
r\sin (\phi)&=\frac{R}{\sqrt{M}z}\sinh\left(\sqrt{M}x\right)\,,\\
r\cos (\phi)&=\frac{R\left(\cosh(\lambda)\sqrt{1-Mz^2}\cosh\left(\sqrt{M}t\right)-\sinh(\lambda)\cosh\left(\sqrt{M} x\right)\right)}{\sqrt{M}z}\,.\label{Map}
\end{aligned}
\end{equation}
Notice the right hand side was already boosted in $\mathbb{R}^{1,1}\subset \mathbb{R}^{2,2}$. We can fix the boost parameter $\lambda$ to be
\be
\tanh\lambda=\sqrt{1-M\epsilon^2}\,.\label{boost}
\ee
This ensures the point particle in global coordinates is at the origin of the AdS${}_3$ space $(r=0)$. The back-reaction in global coordinates is known to be
\be
ds^2=-(r^2+R^2-\mu)d\tilde{\tau}^2+\frac{R^2dr^2}{r^2+R^2-\mu}+r^2d\phi^2\,.
\label{Glob}
\ee
This describes a black hole or a conical defect depending on the ratio $\mu/R^2$. By mapping this metric back to BTZ coordinates using \eqref{Map}, we obtain the time dependent back-reacted solution of Einstein's equations describing a falling massive particle in BTZ. One can check that this solution correctly reproduces the CFT energy density \eqref{eq:local}.

\paragraph{Holographic entanglement entropy :} To compute the holographic entanglement entropy, we use the general result for the covariant entanglement entropy in metric \eqref{Glob} \cite{NNT}
\be
S_A=\frac{c}{6}\left[\log\left(r^{(1)}_\infty\cdot r^{(2)}_{\infty}\right)+\log\frac{2\cos\left(|\Delta \tilde{\tau}_{\infty}|\frac{\sqrt{R^2-\mu}}{R}\right)-2\cos\left(|\Delta \phi_{\infty}|\frac{\sqrt{R^2-\mu}}{R}\right)}{R^2-\mu}\right]
\ee
where $\Delta \phi_{\infty}=\phi^{(2)}_{\infty}-\phi^{(1)}_{\infty}$, $0<|\Delta \phi_\infty|<\pi$,  and $\Delta \tilde{\tau}_{\infty}=\tilde{\tau}^{(2)}_{\infty}-\tilde{\tau}^{(1)}_{\infty}$ describe the global end-points of the geodesic solving the extremal surface equation. Thus, all we have to do is to map these end-points, using the conformal map \eqref{Map}, into their BTZ coordinate counterparts
$(x^{(1)}_\infty,x^{(2)}_\infty)$ at $z_\infty$ and time $t$. After using \eqref{boost} and substituting $\sqrt{M}=2\pi/\beta$, the radial coordinate satisfies
\bea
r^{(i)}_\infty=\frac{R\beta^2}{4\pi^2\epsilon z_\infty}\sqrt{\left(\frac{2\pi\epsilon}{\beta}\right)^2\sinh^2\left(\frac{2\pi x^{(i)}_\infty}{\beta}\right)+\left(\cosh\left(\frac{2\pi t}{\beta}\right)-\sqrt{1-\left(\frac{2\pi\epsilon}{\beta}\right)^2}\cosh\left(\frac{2\pi x^{(i)}_\infty}{\beta}\right)\right)^2}\,,\nn\\
\eea
whereas the boundary coordinates satisfy
\bea
\tan\left(\tilde{\tau}^{(i)}\right)&=&\frac{2\pi\epsilon}{\beta}\frac{\sinh\left(\frac{2\pi t}{\beta}\right)}{\cosh\left(\frac{2\pi x^{(i)}_\infty}{\beta}\right)-\sqrt{1-\left(\frac{2\pi\epsilon}{\beta}\right)^2}\cosh\left(\frac{2\pi t}{\beta}\right)}\nn\\
\tan\left(\phi^{(i)}_\infty\right)&=&\frac{2\pi\epsilon}{\beta}\frac{\sinh\left(\frac{2\pi x^{(i)}_\infty}{\beta}\right)}{\cosh\left(\frac{2\pi t}{\beta}\right)-\sqrt{1-\left(\frac{2\pi\epsilon}{\beta}\right)^2}\cosh\left(\frac{2\pi x^{(i)}_\infty}{\beta}\right)}
\eea
To compare with our CFT computations, we consider $(x^{(1)}_\infty,x^{(2)}_\infty)=(0,L)$. $L$ stands for the size of the subsystem $A$. Since the latter was infinite, we shall be interested in the large $L$ limit, where $L\gg \beta$, keeping the dominant $L$ dependence as a regularised subsystem length.

Our calculation of holographic entanglement entropy shows that at early time $\epsilon\ll t\ll \beta$, $S_A$ grows logarithmically with time $t$ as
\be
S_A\simeq \f{c}{3}\log \left[\f{\beta}{z_{\infty}}e^{\f{\pi L}{\beta}}\right]
+\f{c}{6}\log \f{t}{4\pi^2\ep}+\f{c}{6}\log \left[\f{1}{\s{1-\f{24\Delta_O}{c}}}\sin\left(\pi\s{1-\f{24\Delta_O}{c}}\right)\right].
\label{earlyHEE}
\ee
On the other hand, at late time $t\gg\beta$, it saturates to the value
\be
S_A\simeq \f{c}{3}\log \left[\f{\beta}{z_{\infty}}e^{\f{\pi L}{\beta}}\right]
+\f{c}{6}\log \f{\beta}{8\pi^3\ep}+\f{c}{6}\log \left[\f{1}{\s{1-\f{24\Delta_O}{c}}}\sin\left(\pi\s{1-\f{24\Delta_O}{c}}\right)\right].
\label{lateHEE}
\ee
The infinitesimal parameter $z_{\infty}$ represents the UV cut off of the $z$ coordinate in AdS space.

\begin{figure}[h!]
\centering
\includegraphics[scale=0.8]{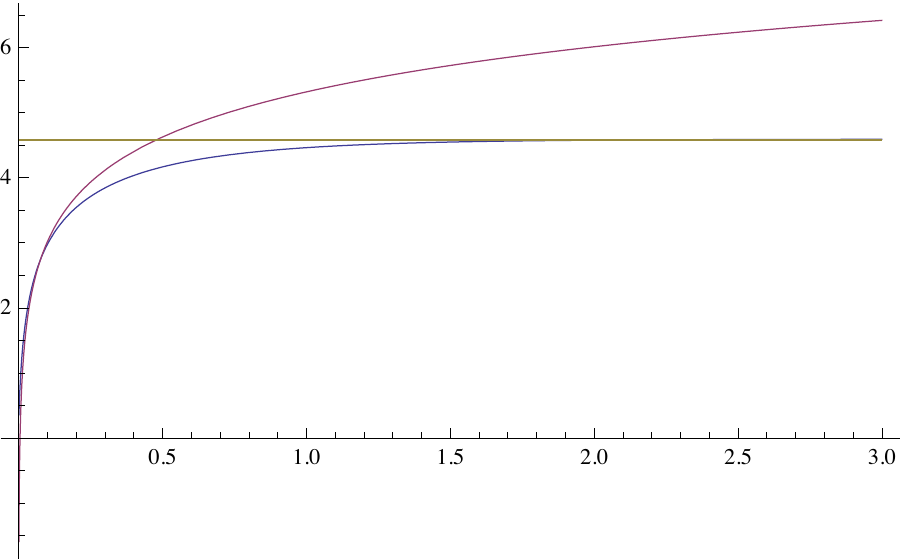}
\caption{The plot of time evolution of holographic entanglement entropy. It shows
$S_A(t)-S_A(0)$ for the infinitely large subsystem $A$ (i.e. a semi half line) as a function of $t$. Blue: full HEE, Red: early time \eqref{earlyHEE}, Yellow: late time \eqref{lateHEE}. We set $R=4G_N=1,\ep=0.001,\beta=3$ and $M=0.1$.}
\label{fig:HEE}
\end{figure}

The first right-hand side terms in \eqref{earlyHEE} and \eqref{lateHEE} coincide with the known result of entanglement entropy for finite temperature CFTs (without any local excitations). Thus, our localized excitations are responsible for the variations on top of the UV answer, as described by the subsequent terms. These variations behave
\begin{itemize}
\item like $\sim \f{c}{6}\log \f{t}{\ep}$ at early time $t\ll \beta$, reproducing the vanishing temperature behaviour found in \cite{NNT},
\item like $\sim \f{c}{6}\log \f{\beta}{\ep}$ at late time $t\gg \beta$.
\end{itemize}
This last behaviour is our new result peculiar to finite temperature. We interpret the saturation to a finite value as due to the fact that for the time $t\gg\beta$ the falling particle is almost stopped at the black hole horizon in the gravity dual.
We plotted the time evolution for specific values of parameters in Fig.\ref{fig:HEE}.\\
Note also that for $\Delta_O$ larger than $c/24$ (the deficit angle geometry becomes the BTZ black hole) the constant terms in \eqref{earlyHEE} and \eqref{lateHEE} are expressed in terms of the effective temperature (see appendix \ref{prs}). It would be interesting to clarify the physical meaning of these universal terms at large c. 

%%%%%%%%%%%%%%%%%%%%%%%%%%%%%%%%%%%%%%%%%%%%%%%%%%%%
%%%%%%%%%%%%%%%%%%%%%%%%%%%%%%%%%%%%%%%%%%%%%%%%%%%%
\section{Mutual information in the thermofield double}
\label{MutualInformation}
%%%%%%%%%%%%%%%%%%%%%%%%%%%%%%%%%%%%%%%%%%%%%%%%%%%%
%%%%%%%%%%%%%%%%%%%%%%%%%%%%%%%%%%%%%%%%%%%%%%%%%%%%

In section~\ref{sec:thermal}, we developed the formalism introduced in \cite{Nozaki:2014hna,He:2014mwa,Caputa:2014vaa} to compute Renyi entanglement entropies in 2d CFTs whose thermal states are locally excited by a primary operator. Our goal in this section is to extend these tools for the set-up described in section~\ref{sec:bulk} involving the thermofield double (TFD) state \eqref{eq:entangled}. Below we will develop calculations of the mutual information in \cite{Mor,HaMa} and compute the mutual information in the thermofield double formalism for our locally excited states.

The main observation, explained in great detail in the appendix of \cite{Mor}, is that just as {\it single} sided thermal correlation functions, such as \eqref{eq:thermal-corr}, must be computed on a {\it single} cylinder with periodicity $\tau \sim \tau + \beta$, {\it two-sided} correlators, such as \eqref{eq:two-sided}, involve a path integral over a cylinder with the same periodicity $\tau \sim \tau + \beta$, where {\it all} operators $\mathcal{O}_R$ acting on $\mathcal{H}_R$ are inserted at $\tau=i\beta/2$, whereas operators $\mathcal{O}_L$ acting on $\mathcal{H}_L$ are still inserted at $\tau=0$ . This is consistent with the analytic continuation \eqref{eq:anal-cont}.

\begin{figure}[h!]
\centering
\includegraphics[width=8cm]{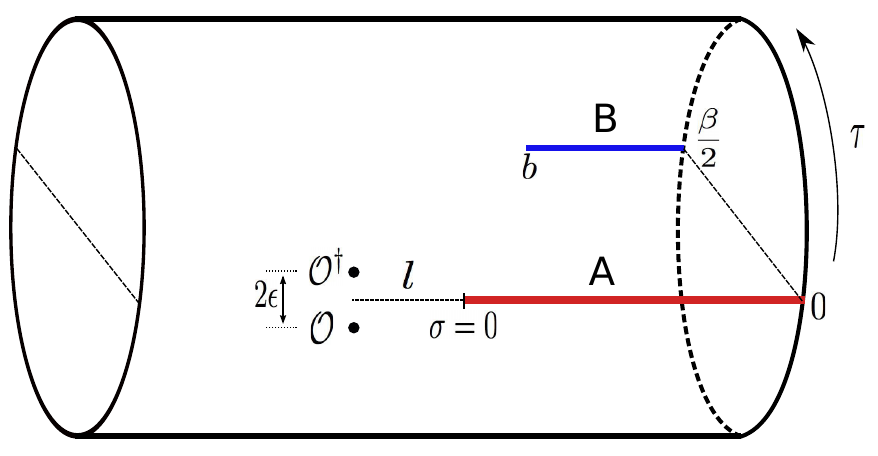}
\caption{Reduced density matrix in the thermofield double state. CFT$_L$ corresponds to $x=i0$ and CFT$_R$ to $x=i\frac{\beta}{2}$ \label{fig:Cyl2}}
\end{figure}

%%%%%%%%%%%%%%
\paragraph{Our set-up :} We consider the thermofield double state \eqref{eq:entangled} and two semi-infinite intervals: $A=[ 0,\infty]$ in the left CFT$_L$ and $B=[b,\infty]$ in the right CFT$_R$. We perturb the TFD by the insertion of a local primary operator $\mathcal{O}_L$ acting on CFT$_L$ at $x=-l\,, t=0$ and follow the real time evolution of the change in the mutual information defined in terms of the growths of Renyi entanglement entropies introduced in the previous section as
\be
\Delta I^{(2)}_{A:B}=\Delta S^{(2)}_A+\Delta S^{(2)}_B-\Delta S^{(2)}_{A\cup B}.
\ee
By definition \eqref{Renyin}, the second Renyi entanglement entropies $\Delta S^{(2)}_X$, where $X\in\{A,B,A\cup B\}$, require the computation of $4$-point correlators on two cylinders glued along the corresponding cuts. At $t=0$, the interval A corresponds to a cut at $\tau=0$, whereas the interval B corresponds to a cut at $\tau=i\beta/2$. The operator insertions in the first cylinder remain equal to $x_1$ and $x_2$, i.e. $-l\pm i\epsilon$ at $t=0$, whereas the insertion points on the second cylinder are
\be
x_3=x_1+i\beta,\qquad x_4=x_2+i\beta
\ee
and similarly for $\bar{x}_{3}$ and $\bar{x}_4$. Figure~\ref{fig:Cyl2} summarises the entire construction.

As reviewed in section~\ref{sec:bulk}, there are two different natural time evolutions to consider. Using the hamiltonian
$H_R-H_L$, the time dependent density matrix is
\begin{equation}
\begin{aligned}
\rho(t)&=\mathcal{N}e^{-i(H_R-H_L)t}e^{-\epsilon H_L}\mathcal{O}_L(-l)\ket{\Psi_\beta}\bra{\Psi_\beta}\mathcal{O}^\dagger_L(-l)e^{-\epsilon H_L}e^{i(H_R-H_L)t}\\
&\equiv\mathcal{N}\mathcal{O}_L(x_2,\bar{x}_2)\ketbra{\Psi_\beta}\mathcal{O}^\dagger_L(x_1,\bar{x}_1)\,.
\end{aligned}
\end{equation}
On the other hand, the hamiltonian $H_R+H_L$ gives rise to the density matrix
\begin{equation}
\begin{aligned}
  \rho(t)&=\mathcal{N}e^{-i(H_L+H_R)t}e^{-\epsilon H_L}\mathcal{O}_L(-l)\ket{\Psi_\beta}\bra{\Psi_\beta}\mathcal{O}^\dagger_L(-l)e^{-\epsilon H_L}e^{i(H_L+H_R)t}\\
  &=\mathcal{N}\mathcal{O}_L(x_2,\bar{x}_2)e^{-i(H_L+H_R)t}
  \ketbra{\Psi_\beta}e^{i(H_L+H_R)t}\mathcal{O}^\dagger_L(x_1,\bar{x}_1).
\end{aligned}
\end{equation}
Technically, the main difference is that under $H_R+H_L$ evolution, both the local operator and the state $\ket{\Psi_\beta}$ evolve non-trivially, whereas for $H_R-H_L$ only $\mathcal{O}_L$ does, since  $\ket{\Psi_\beta}$ is boost invariant.

To sum up, for both hamiltonians the positions of the operators depend on time but for $H_R-H_L$ the position of the cuts is fixed, whereas for $H_R+H_L$, the semi-infinite interval $A$ starts at fixed $(x_A,\bar{x}_A)=(0,0)$ but the starting point of the semi-infinite cut B $(x_B,\bar{x}_B)=(b+i\beta/2+2t,b-i\beta/2-2t)$ depends time (see \cite{HaMa}). We will now analyze the real time evolution of $\Delta I^{(2)}_{A:B}$ using the appropriate conformal maps. For each of the hamiltonians we will present the results for the free boson CFT and a general 2d CFT at large central charge. A physical picture behind all the results for mutual information will be given in section \ref{PhysP}.

%%%%%%%%%%%%%%%%%%%%%%%%%%%%%%%%%%%%%%%%%%%%%%%%%%%%
\subsection{$H_R-H_L$ evolution}
%%%%%%%%%%%%%%%%%%%%%%%%%%%%%%%%%%%%%%%%%%%%%%%%%%%%

The calculation of $\Delta S^{(2)}_X$ involves a 4-pt function. We follow the same strategy as in section~\ref{sec:thermal}. That is, we compute the relevant 4-pt function by mapping our two glued cylinders to the plane. These maps are given below for the different $X$ :
\begin{equation}
\begin{aligned}
  z(x)&=\sqrt{e^{\frac{2\pi}{\beta}x}-1}\,, \quad \text{for} \quad  X=A\\
  z(x)&=\sqrt{e^{\frac{2\pi}{\beta}x}+e^{\frac{2\pi}{\beta}b}}\,, \quad \text{for} \quad X=B\\
  z(x)&=\sqrt{\frac{e^{\frac{2\pi}{\beta}x}-1}{e^{\frac{2\pi}{\beta}x}+e^{\frac{2\pi b}{\beta}}}}\,, \quad \text{for} \quad X=A\cup B\,.
\label{MapsLmR}
\end{aligned}
\end{equation}
The same maps apply for $\bar{z}(\bar{x})$. These were obtained as the map \eqref{MapSI} : they are the composition of a map from the cylinder to the plane followed by a uniformization map dealing with the existing cuts.

Since all maps \eqref{MapsLmR} satisfy $z_3=-z_1$ and $z_4=-z_2$, all $\Delta S^{(2)}_X$ will still satisfy
\be
\Delta S^{(2)}_X=-\log\left(\left|z_X(1-z_X)\right|^{4\Delta_O}G(z_X,\bar{z}_X)\right)\,, \quad \quad X=\{A,B,A\cup B\}\,,
\label{eq:DSmain}
\ee
as already derived in section~\ref{sec:thermal}. The explicit form of the different cross-ratios $z_X,\,\bar{z}_X$ determining the behaviour of these Renyi entanglement entropies is presented in equations \eqref{CRSI}, \eqref{CRBm} and \eqref{CRABm}. In the following, we discuss the small $\epsilon$ expansion for different times $t$ in different CFTs.

%%%%%%%%%%%%%%%%
\paragraph{Free boson :} Using \eqref{GFS} in \eqref{eq:DSmain} with three different cross-ratios, the change in mutual information equals
\be
\Delta I^{(2)}_{A:B}=\log\left(2\frac{1+|z_{A\cup B}|+|1-z_{A\cup B}|}{\left(1+|z_{A}|+|1-z_{A}|\right)\left(1+|z_{B}|+|1-z_{B}|\right)}\right)\,.
\ee

\begin{figure}[b!]	
  \centering
  \includegraphics[width=10cm]{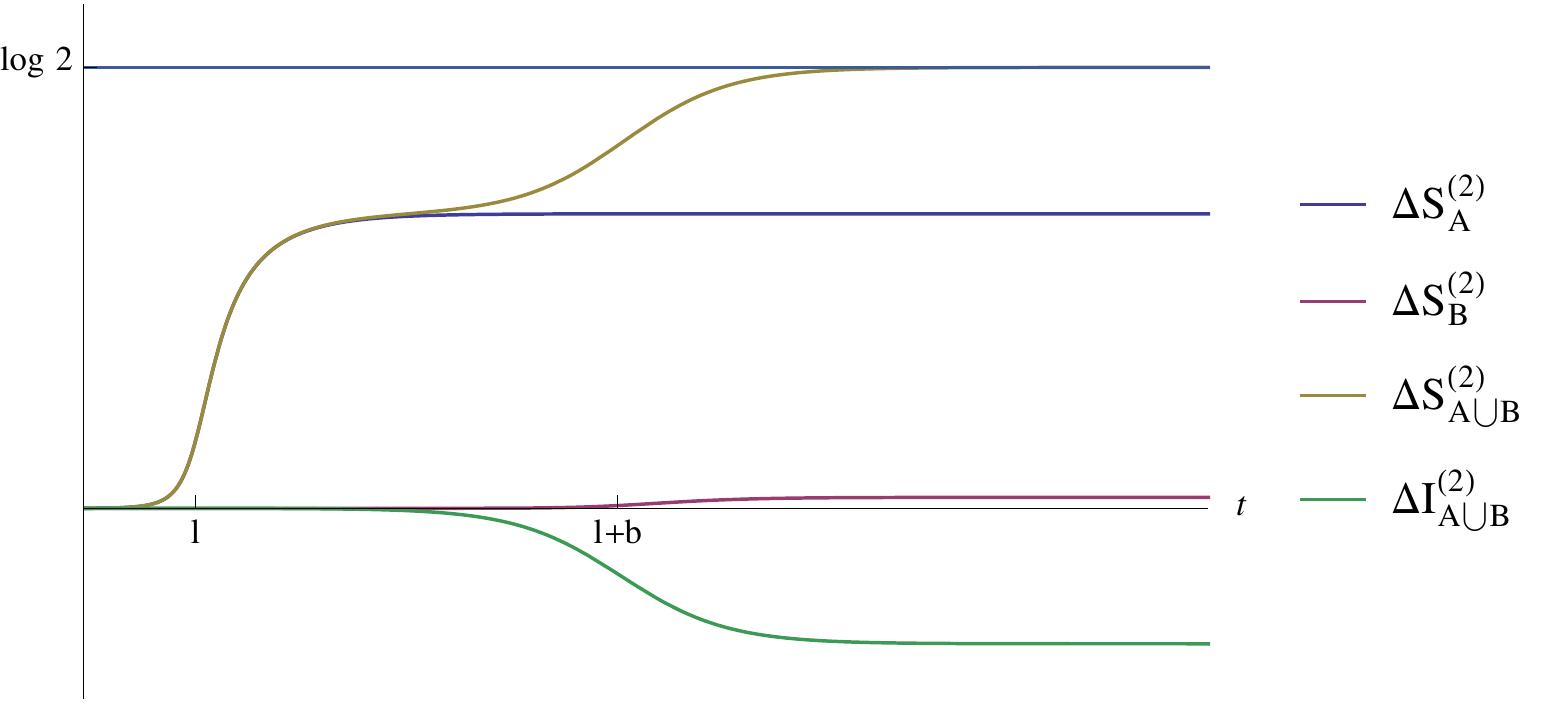}
  \caption{Growth of the Renyi entanglement entropies for free scalar in TDF. Parameters $\epsilon/\beta=1/12$ and $b\neq 0$.}\label{FSMIb}
\end{figure}
Figure \ref{FSMIb} shows the three growths $\Delta S^{(2)}_X$ and the change in the mutual information $\Delta I^{(2)}_{A:B}$.
For small $\frac{\epsilon}{\beta}$, the entropy $\Delta S^{(2)}_B$ is negligible, i.e. $O(\epsilon^2)$, for all times. Thus, the change in the mutual information reduces to
\begin{equation}
  \Delta I^{(2)}_{A:B}=\Delta S^{(2)}_A -\Delta S^{(2)}_{A\cup B} + O(\epsilon^2)\,.
\label{eq:approx}
\end{equation}
Both $\Delta S^{(2)}_A$ and $\Delta S^{(2)}_{A\cup B}$ approach $\log 2-\frac{\pi\epsilon}{\beta}$ in the time interval $l<t<l+b$, whereas
for $t>l+b$, the growth of the Renyi entanglement entropy of the union increases to the maximal value
\be
\Delta S^{(2)}_{A\cup B}\to \log 2\,, \quad t > l + b\,.
\ee
As a result, the change in the mutual information is mostly zero up to the time $t\simeq l+b$ and saturates at the negative value
\be
\Delta I^{(2)}_{A:B}\to -\frac{\pi\epsilon}{\beta}+O(\epsilon^2)\,,
\ee
soon afterwards.

\begin{figure}[h!]	
  \centering
  \includegraphics[width=10cm]{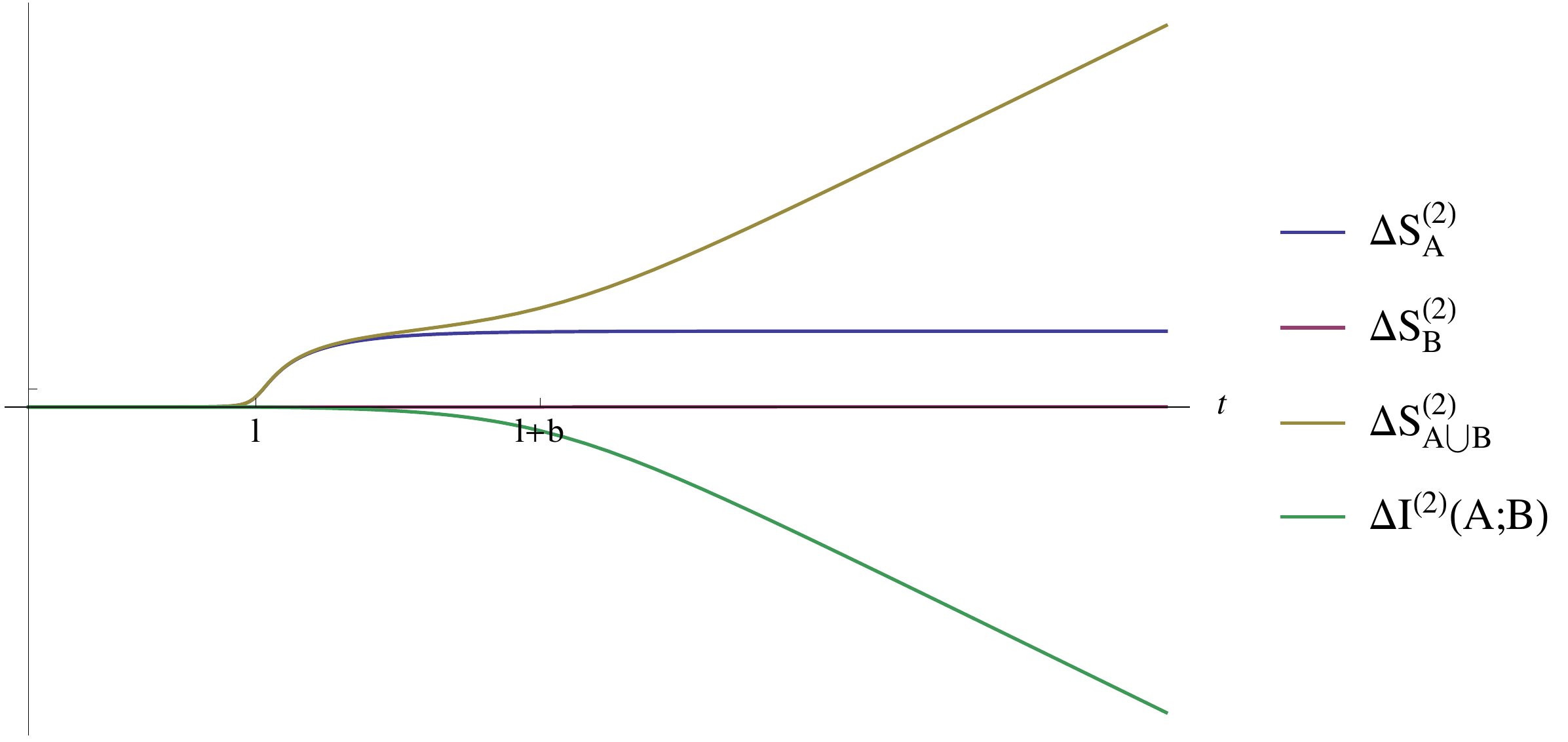}
  \caption{Growths on the Renyi entanglement entropies and the change in the mutual information at large c for $\frac{\epsilon}{\beta}=\frac{1}{30}$}\label{Renyi}
\end{figure}

\paragraph{Large $c$ general 2d CFT :} Using the large c behavior of the conformal blocks \eqref{Fateev} and the form of the growth of the Renyi entanglement entropies \eqref{LargeCDS}. the change in mutual information is
\be
\Delta I^{(2)}_{A:B}\simeq 4\Delta_O\log\left(\frac{|1-z_{A\cup B}|}{|1-z_A||1-z_B|}\right)\,.
\label{MILc}
\ee
In figure~\ref{Renyi}, we plot the three growths $\Delta S^{(2)}_X$ and the change in the mutual information.
Notice $\Delta S^{(2)}_B$ is zero for all times. Since $\Delta S^{(2)}_A,\,\Delta S^{(2)}_{A\cup B} \to 4\Delta_O\log\left(\frac{\beta}{\pi\epsilon}\right)$ for $l < t < l +b $ (as in \eqref{RELc} ), $\Delta I^{(2)}_{A:B}$ still vanishes in this interval. When $t > l +b$, $\Delta S^{(2)}_{A\cup B}$ grows linearly with time. This results in the linear decrease of the change in the mutual information from $0$ to $-\infty$ :
\be
\Delta I^{(2)}_{A:B}\simeq -\frac{8\pi \Delta_O t}{\beta}\,.
\ee
Interestingly, since $\Delta S^{(2)}_B$ is negligible, all $\epsilon$ dependence cancels at this order.

%%%%%%%%%%%%%%%%%%%%%%%%%%%%%%%%%%%%%%%%%%%%%%%%%%%%
\subsection{$H_R+H_L$ evolution}
%%%%%%%%%%%%%%%%%%%%%%%%%%%%%%%%%%%%%%%%%%%%%%%%%%%%

We follow the same strategy as in previous sections and compute the relevant 4-pt function using appropriate conformal maps. When time evolution is generated by $H_R+H_L$, these are
\begin{equation}
\begin{aligned}
  z(x)&=\sqrt{e^{\frac{2\pi}{\beta}x}-1},\qquad\qquad\quad \bar{z}(\bar{x})=\sqrt{e^{\frac{2\pi}{\beta}\bar{x}}-1}\,, \quad \text{for} \quad  X=A \\
  z(x)&=\sqrt{e^{\frac{2\pi}{\beta}x}+e^{\frac{2\pi}{\beta}(b+2t)}},\qquad \bar{z}(\bar{x})= \sqrt{e^{\frac{2\pi}{\beta}\bar{x}}+e^{\frac{2\pi}{\beta}(b-2t)}}\,, \quad \text{for} \quad  X=B \\
  z(x)&=\sqrt{\frac{e^{\frac{2\pi}{\beta}x}-1}
  {e^{\frac{2\pi}{\beta}x}+e^{\frac{2\pi}{\beta}(b+2t)}}},\qquad \bar{z}(\bar{x})=\sqrt{\frac{e^{\frac{2\pi}{\beta}\bar{x}}-1}
  {e^{\frac{2\pi}{\beta}\bar{x}}+e^{\frac{2\pi}{\beta}(b-2t)}}}\,, \quad \text{for} \quad  X=A\cup B
\label{MapsLpR}
\end{aligned}
\end{equation}
Notice these maps encode the time evolution of the cuts inserted at $\tau = i\frac{\beta}{2}$ through the shifts $b\to b+2t$ for functions of $x$ and $b\to b-2t$ for functions of $\bar{x}$, when compared with the maps \eqref{MapsLmR}. But they still satisfy $z_3=-z_1$ and $z_4=-z_2$. This means that the individual growths $\Delta S^{(2)}_X$  will be computed by \eqref{eq:DSmain}. This requires the computation of the new cross-ratios $(z_B,\bar{z}_B)$ and $(z_{A\cup B},\bar{z}_{A\cup B})$, given in \eqref{CRBp} and  \eqref{CRABp} respectively.

\begin{figure}[h!]	
  \centering
  \includegraphics[width=10cm]{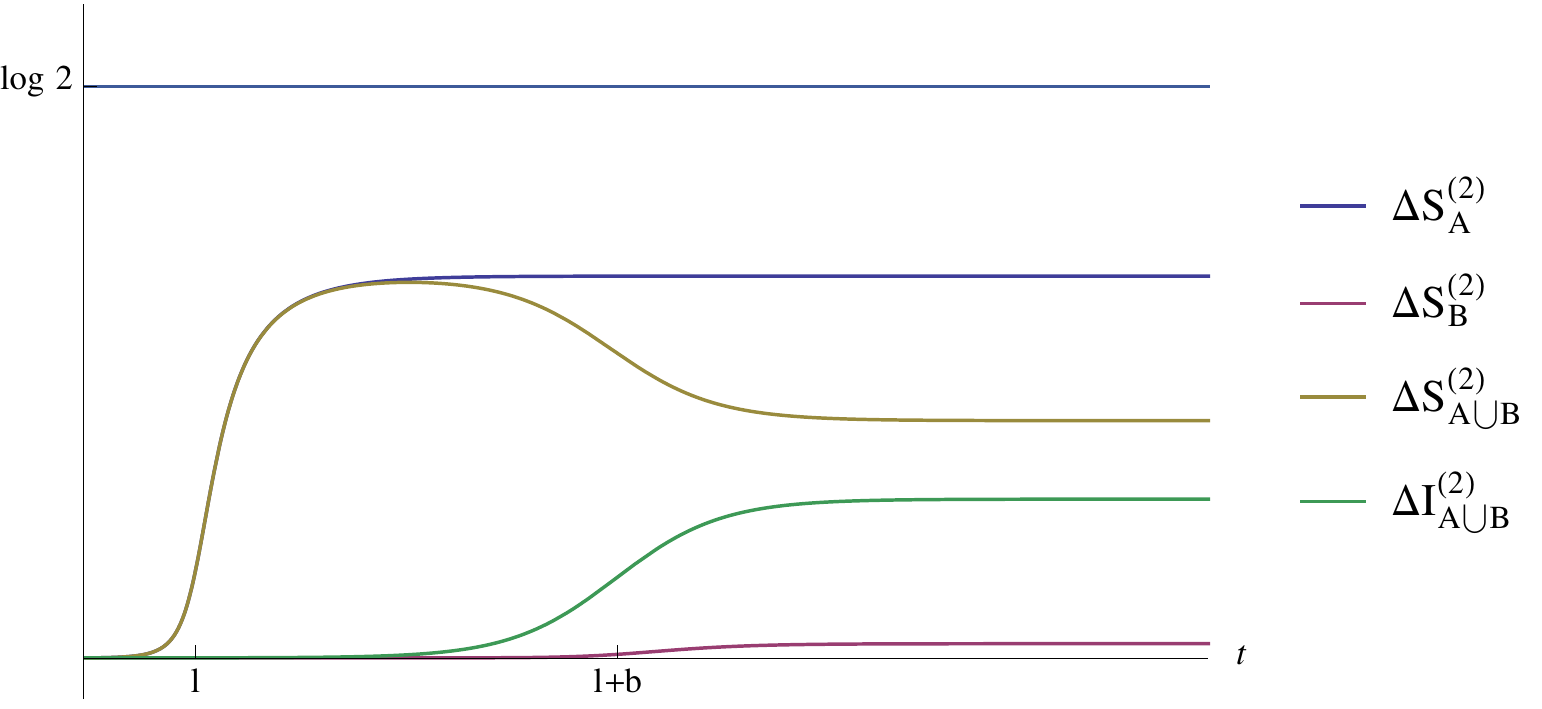}
  \caption{Growths in the Renyi entanglement entropies and the change in the mutual information for the free boson for $\epsilon/\beta=1/12$ and $b\neq 0$ .}\label{FSMIpb}
\end{figure}

\paragraph{Free boson :} There is no change in any of the $\Delta S_X^{(2)}$ up to $t=l+b$ when compared to $H_R- H_L$ evolution
 (see figure~\ref{FSMIpb}). However, for $t>l+b$, $\Delta S^{(2)}_{A\cup B}$ approaches
\be
\Delta S^{(2)}_{A\cup B}\to \log(2)-\frac{2\pi\epsilon}{\beta}\,.
\ee

Notice how the maximal value of $\log 2$ is now decreased twice as much as for $\Delta S^{(2)}_A$. This implies that for $H_R+H_L$ evolution, at late time and small $\epsilon$, the change in the mutual information increases by the same amount as its counterpart decreases for the hamiltonian $H_R-H_L$
\be
\Delta I^{(2)}_{A:B}\to \frac{\pi\epsilon}{\beta}+O(\epsilon^2)\,.
\ee

\begin{figure}[h!]	
  \centering
  \includegraphics[width=9.5cm]{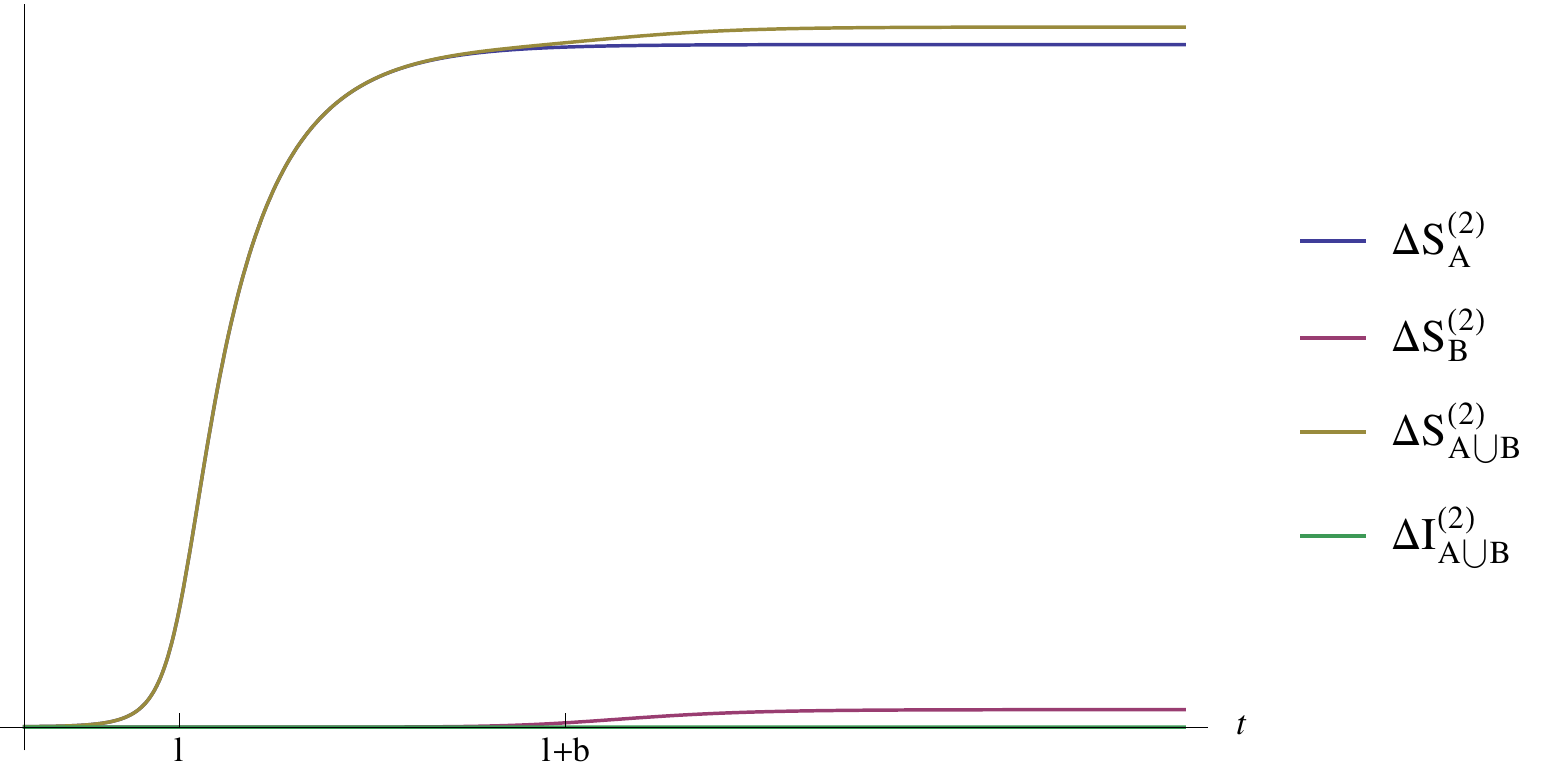}
  \caption{Growths of Renyi entanglement entropies and mutual information for $H_R+H_L$ evolution, $\epsilon/\beta=\frac{1}{8}$}\label{MIPall}
\end{figure}

\paragraph{Large $c$ generic 2d CFT :} The variation in mutual information is still given by \eqref{MILc}, but using the appropriate cross-ratios dealing with the evolution generated by $H_R+H_L$. Results are plotted in figure~\ref{MIPall}. Surprisingly, there is no change in the mutual information for all times in the small $\epsilon$ limit. Technically, the reason this happens is because when $t>l+b$, the difference between $\Delta S^{(2)}_{A\cup B}$ and $\Delta S^{(2)}_A$ is order $O(\epsilon^2)$. This is the same order as the neglected contribution from  $\Delta S^{(2)}_B$. Thus, $\Delta I^{(2)}_{A\cup B}$ vanishes, at this order, for late times.

\begin{figure}[h!]	
  \centering
  \includegraphics[width=9cm]{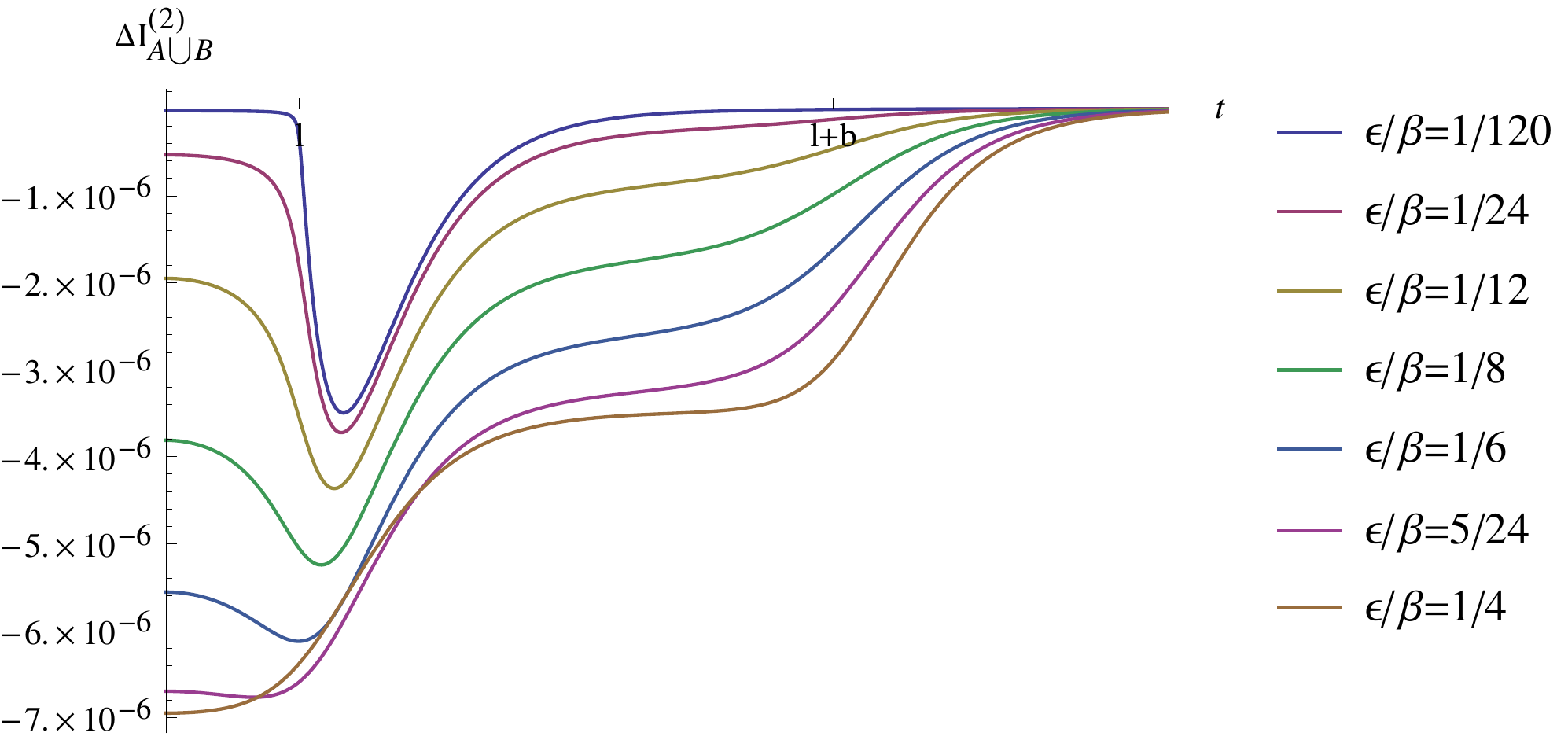}
  \caption{Mutual information for $H_R+H_L$ evolution, $b=4l$}\label{MILCpb}
\end{figure}

Since we know the exact dependence of the cross-ratios on all our parameters, we can check how and when the growths $\Delta S^{(2)}_X$ as well as the change in the mutual information $\Delta I^{(2)}_{A:B}$ differ. Under $H_R+H_L$ evolution, it turns out there exists a small difference in the regime $0< t <l+b$. We plot $\Delta I^{(2)}_{A:B}$ in this region for several values of $\frac{\epsilon}{\beta}$ in
figure~\ref{MILCpb}.
\\
%%%%%%%%%%%%%%%%%%%%%%%%%%%%%%%%%%%%%%%%%%%%%%%%%%%%
\subsection{Possible Physical interpretation}\label{PhysP}
%%%%%%%%%%%%%%%%%%%%%%%%%%%%%%%%%%%%%%%%%%%%%%%%%%%%

\begin{figure}[h!]	
  \centering
  \includegraphics[width=10cm]{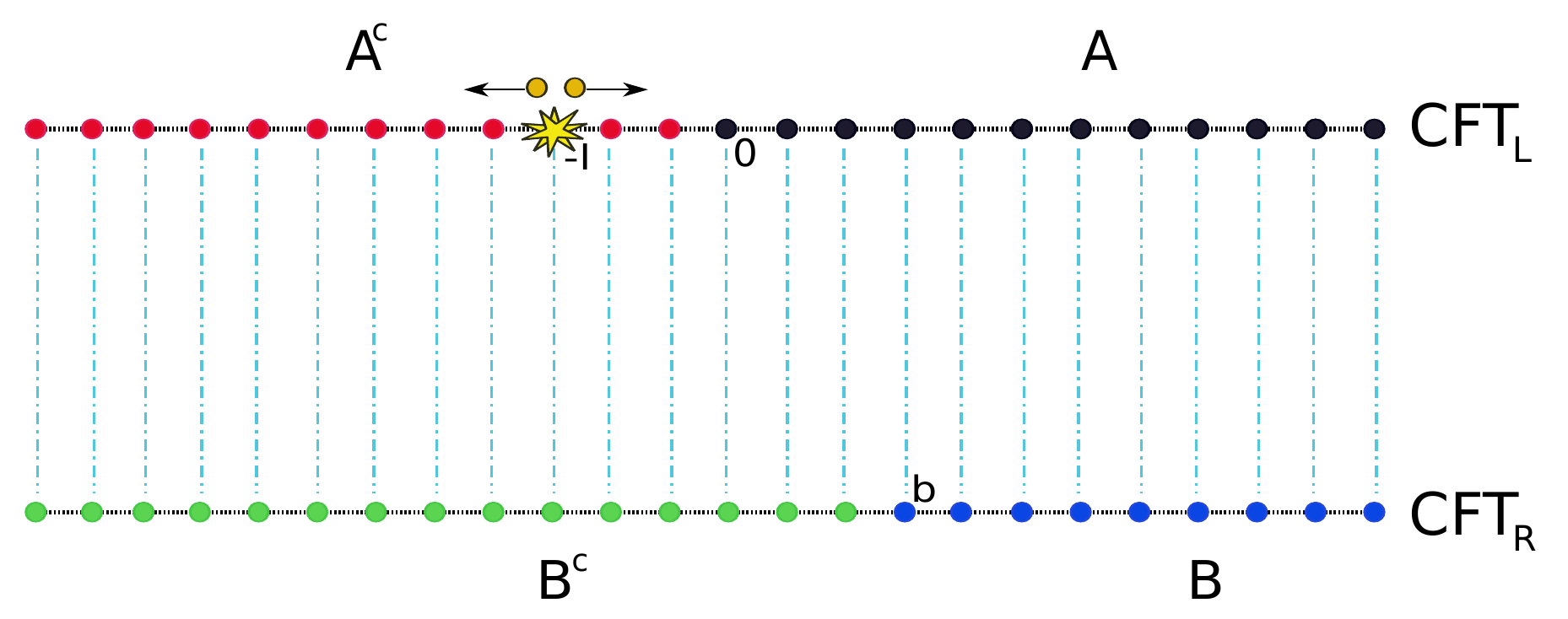}
  \caption{The structure of entanglement for the TFD at $t=0$. The upper and lower chain describe
  a lattice analogue of CFT$_L$ and CFT$_{R}$, respectively. In CFT$_L$, we insert the local operator at $x=-l$, which creates entangled pair described by the yellow dots. They propagate at the speed of light in the opposite directions.}\label{T0}
\end{figure}

In this subsection, we would like to provide a possible physical explanation for our results on the growth of Renyi entanglement entropies and mutual information in terms of the propagating entangled pairs created by the local operator $\mathcal{O}$ in TFD (fig.\ref{T0}). Our energy density calculations in section~\ref{sec:density} suggest that, in the limit $\frac{\epsilon}{\beta}\ll 1$, the positions of this pair at time $t$ are peaked at $x=\pm t - l$. In the following, we refer to $A^c$ and $B^c$ as the complementary regions to $A$ and $B$ in the entire CFT$_L$ and CFT$_R$ spaces, respectively.

In all our examples, $\Delta S^{(2)}_{B}$ is always negligible in the approximation $\f{\ep}{\beta}\ll 1$. This is natural because it is highly non-trivial (but not impossible) for the excitation in CFT$_L$ to affect the entanglement entropy in CFT$_R$. Keeping this in mind, we focus our attention on the behavior of $\Delta S^{(2)}_A$ and $\Delta S^{(2)}_{A\cup B}$ below. The mutual information is given by the difference between these two quantities, as in \eqref{eq:approx}.

%%%%%%%%%%%%%%%%%%%%%
\paragraph{Free Scalar with $H=H_R-H_L$ :} When $0<t<l$, causality ensures the entangled pair created by the local operator $\mathcal{O}_L$ remains in $A^c$. Thus, $\Delta S^{(2)}_X$ is trivial. When $l<t<l+b$, one part of each entangled pair enters into $A$. Since the original thermofield double state \eqref{eq:entangled} is highly entangled from the perspective of ${\cal H}_L$ and ${\cal H}_R$, the addition of this extra entanglement is expected to break the original entanglement due to the monogamy property. By regarding CFT$_L$ and CFT$_R$ as our laboratory and thermal bath, this is a basic example of decoherence. Since thermal entropy density is proportional to $1/\beta$, the reduction in entanglement between $A$ and $B^c$ is expected to be of the order $\epsilon/\beta$ to take into account the width of the local perturbation. Thus, both $\Delta S^{(2)}_A$ and $\Delta S^{(2)}_{A\cup B}$ are reduced by $O(\ep/\beta)$. This explains our CFT result $\Delta S^{(2)}_A=\Delta S^{(2)}_{A\cup B}=\log 2-\f{\pi\ep}{\beta}$.

At later times, when $t>l+b$, the excitation is propagating through points in $A$ which are thermally entangled with $B$. Thus, the entanglement breaking mechanism described for $l < t < l + b$ does not apply for $\Delta S^{(2)}_{A\cup B}$ (since this would require the entangled pair to involve $A$ and $B^c$ or $A^c$ and $B$). Thus in this time region, $\Delta S^{(2)}_A$ is still decreased to $\log 2-\f{\pi\ep}{\beta}$, whereas $\Delta S^{(2)}_{A\cup B}=\log 2$, in agreement with our CFT calculations.

\begin{figure}[h!]	
  \centering
  \includegraphics[width=6cm]{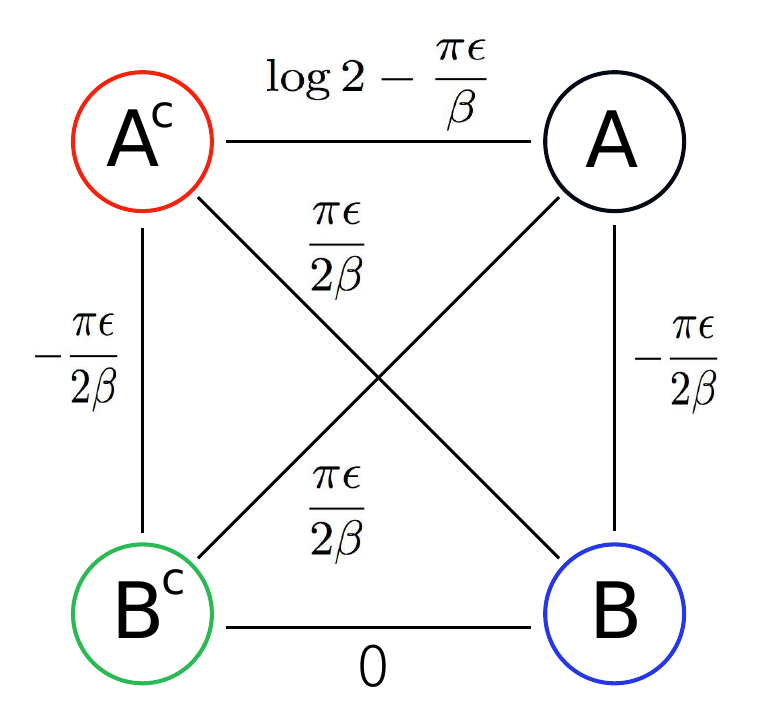}
  \caption{Late time ($t>l+b$) entangled structure for the free boson TFD with $H=H_R-H_L$}\label{ESm}
\end{figure}

This heuristic picture is summarized in figure~\ref{ESm}, in terms of a composite system of 4 subsystems $A,\, A^c,\,B,\,B^c$ and the amount of entanglement gained or reduced when considering different pairings. We assumed that the entanglement between $B$ and $B^c$ is vanishing to draw this picture.

\begin{figure}[h!]	
  \centering
  \includegraphics[width=6cm]{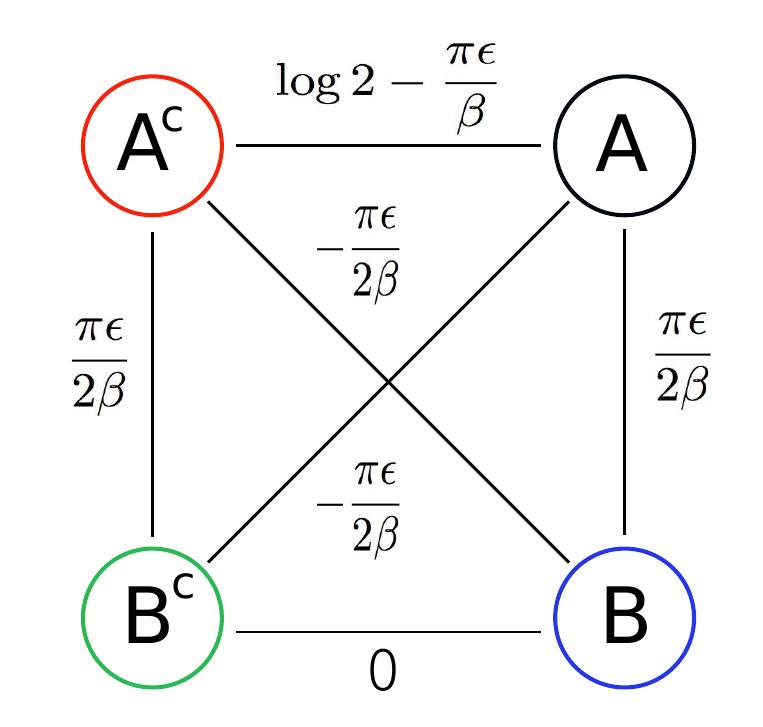}
  \caption{Late time ($t>l+b$) entangled structure for the free boson TFD with $H=H_R+H_L$}\label{ESp}
\end{figure}

%%%%%%%%%%%%%%%%%%%%%
\paragraph{Free Scalar with $H=H_R+H_L$ :} As previously stressed, the evolution under $H_R+H_L$ makes the location of the cut $B$ move with time. In other words, at the time $t$ generated by
$H_R+H_L$, the thermofield double entanglement leads to the entanglement between a point $x$ in CFT$_L$ with $x\pm 2t$ in CFT$_R$. Our discussion and interpretation do not change for $l < t < l+b$ since in this time interval the variations in the amount of entanglement are due to entanglement between $A$ and $B^c$.  When $t>l+b$, $\Delta S^{(2)}_A$ remains unchanged but $\Delta S^{(2)}_{A\cup B}$ is further decreased due to the extra breaking of entanglement between $A^c$ and $B$, qualitatively reproducing our CFT results $\Delta S^{(2)}_A=\log 2-\f{\pi\ep}{\beta}$ and
$\Delta S^{(2)}_{A\cup B}=\log 2-\f{2\pi\ep}{\beta}$. This is again heuristically represented in figure~\ref{ESp}.

%%%%%%%%%%%%%%%%%%%%%
\paragraph{Large c with $H=H_R-H_L$ :} Even though the large $c$ limit may typically correspond to a strong interacting CFT limit, we can still provide a qualitative interpretation for our results using similar arguments to the ones in the free scalar case. The main difference encountered in the large c limit is that $\Delta S^{(2)}_{A\cup B}\sim -\Delta I^{(2)}_{A:B}\sim \f{8\pi\Delta}{\beta}t$ at late times $(t > l+b)$. This linear $t$ behavior may be interpreted by the effect of breaking of original entangled pairs that remains at later time as opposed to the free scalar theory. This can be due to the fact that we are considering the large $c$ strongly interacting CFTs and thus the entanglement breaking effect of the local operator is strong enough to last forever once it happens.

%%%%%%%%%%%%%%%%%%%%%
\paragraph{Large c with $H=H_R+H_L$ :} In this case, we find trivial results for the change in the mutual information at any time. The reason why $\Delta S^{(2)}_{A\cup B}$ is suppressed is because the local excitation does not break the original entanglement between $A$ and $B$ but it does the one between $A$ (or $B$) and $B^c$ (or $A^c$).

%%%%%%%%%%%%%%%%%%%%%%%%%%%%%%%%%%%%%%%%%%%%%%%%%%%%
%%%%%%%%%%%%%%%%%%%%%%%%%%%%%%%%%%%%%%%%%%%%%%%%%%%%
\section{Conclusions}
%%%%%%%%%%%%%%%%%%%%%%%%%%%%%%%%%%%%%%%%%%%%%%%%%%%%
%%%%%%%%%%%%%%%%%%%%%%%%%%%%%%%%%%%%%%%%%%%%%%%%%%%%

In this paper, we first studied the growth of (Renyi) entanglement entropy in the presence of locally excited states in two dimensional conformal field theories (2d CFTs) at finite temperature. We define our excited state by acting with a local operator $\mathcal{O}(x)$ with conformal dimension $\Delta_O$ on the CFT thermal state. In all calculations in this paper we took the subsystem $A$ in the
definition of the reduced density matrices $\rho_A$ to be a semi-infinite line.
We found the general formula \eqref{DSmain} for the growth of the second Renyi entanglement entropy $\Delta S^{(2)}_A$ in any 2d CFTs in terms of the four point function of the local operators. The result depends on the temperature $T=1/\beta$ and UV regularization parameter $\ep$ of the local operator. We chose $\ep$ such that the local operator is smeared over a small width $\ep$.

We explicitly evaluated this formula both for massless free field theories and CFTs in the large central charge $c$ limit (assuming $\Delta_O \ll c$). If we are allowed to take strictly $\ep=0$, which is not possible in the latter case, then the result of $\Delta S^{(2)}_A$ does not depend on the temperature $T$. In the former case (i.e. free CFTs), we found that the growth of entropy is reduced by $O(\ep /\beta)$ compared with the zero temperature case. This is interpreted as a decoherence effect, where the original entanglement between the CFT and its thermofield double is reduced because the local excitation (with the size $\ep$) also breaks the original quantum entanglement.   In the large $c$ limit case, the zero temperature result behaves like $\sim \Delta_O \log\f{t}{\ep}$, assuming $\Delta_O \ll c$. At finite temperature, we found that the time evolution is stopped at time $t\sim \beta$ and $\Delta S^{(2)}_A$ gets saturated to a value $\sim \Delta_O\log\f{\beta}{\ep}$.

We calculated the holographic entanglement entropy in a gravity dual of this local operator excitation at finite temperature. Our gravity description is given by a falling massive particle in a BTZ black hole. Since this takes into account the full back-reaction, the result corresponds to the local operator with a very large conformal dimension $\Delta \sim O(c)$, as opposed to our large $c$ limit analysis of 2d CFTs. Despite their different regime of validity, we find it takes time $t\sim \beta$ for the massive particle to approach the horizon, a time scale that seems to agree with our analysis of Renyi entanglement entropy. Furthermore, we analytically evaluate the time evolution of holographic entanglement entropy and found that it is indeed saturated around $t\sim \beta$ to a value $\Delta S_A\sim \f{c}{6}\log \f{\beta}{\ep}$.

In the last part of this paper, we studied further the structure of quantum entanglement by employing the thermofield double (TFD) description of finite temperature CFTs. This has the nice advantage that we can study patterns of entanglement in an easier way as the total system is described by a pure state. We studied the behavior of Renyi entanglement entropy for two different time evolutions in free scalar field theories and large $c$ 2d CFTs. The first evolution is with respect to the time which is the Killing symmetry in the dual eternal AdS black holes, which corresponds to the Hamiltonian $H_R-H_L$. Here $H_L$ and $H_R$ are the Hamiltonians of the two CFTs in the thermofield double. The second evolution is the non-trivial time evolution generated by $H_R+H_L$, which corresponds to the time evolution along so called nice slices of eternal AdS black holes.

We formulated replica method calculations for both choices of the time evolutions and evaluated explicitly the growth of the mutual information between two semi-infinite lines on the two CFTs in the presence of local operator excitations. In the first case $H_R-H_L$, we found that the mutual information decreases in general. This is naturally explained by noting that the entangled pairs created by the local operator break the initial entanglement between the two CFTs originated from the thermofield double construction. In the second choice $H_R+H_L$, we found that the mutual information does not show this decreasing behavior. This can be again understood intuitively. For this choice of time evolution the time $t$ in the first CFT corresponds to the time $-t$ in the second CFT. Therefore the entanglement which arises from the thermofield double construction connect two points in the two CFTs which are separated by $2t$. Thus the breaking of entanglement due to the local operators does not contribute to the mutual information. It will be an interesting future problem to confirm these CFT results from the calculations in some gravity duals.

%%%%%%%%%%%%%%%%%%%%%%%%%%%%%%%
\subsection*{Acknowledgements}
%%%%%%%%%%%%%%%%%%%%%%%%%%%%%%%

We would like to thank Masahiro Nozaki, Tokiro Numasawa and Kento Watanabe for useful discussions and Alvaro Veliz Osorio for comments on the draft. We are also grateful to Tom Hartman for email correspondence. PC is supported by JSPS postdoctoral fellowship for overseas researchers and from the 1st of November by  the Swedish Research Council (VR) grant 2013-4329. TT is supported by JSPS Grant-in-Aid for Scientific Research (B) No.25287058 and by JSPS Grant-in-Aid for Challenging Exploratory Research No.24654057. TT is also supported by World Premier International Research Center Initiative (WPI Initiative) from the Japan Ministry of Education, Culture, Sports, Science and Technology (MEXT). The work of JS and AS was partially supported by the Science and Technology Facilities Council (STFC) [grant number ST/J000329/1]. 

%\newpage
%%%%%%%%%%%%%%%%%%%%%%%%%%%%%%%%%%%%%%%
%%%%%%%%%%%%%%%%%%%%%%%%%%%%%%%%%%%%%%%
\begin{appendix}
%%%%%%%%%%%%%%%%%%%%%%%%%%%%%%%%%%%%%%%
%%%%%%%%%%%%%%%%%%%%%%%%%%%%%%%%%%%%%%%

%%%%%%%%%%%%%%%%%%%%%%%%%%%%%
%%%%%%%%%%%%%%%%%%%%%%%%%%%%%
\section{Primary States as Excited States}\label{prs}
%%%%%%%%%%%%%%%%%%%%%%%%%%%%%
%%%%%%%%%%%%%%%%%%%%%%%%%%%%%

Consider a 2d CFT at zero temperature for simplicity. We should distinguish our locally excited state $\mathcal{O}(x)|0\lb$ from the standard primary state $|\mathcal{O}\lb$ which corresponds to a primary operator $\mathcal{O}$. The latter is defined from the former by taking the far past limit $\tau\to -\infty$ with respect to the Euclidean time $\tau$ and it has a definite conformal dimension $\Delta_O$ as the eigenvalue of the dilatational operator. On the other hand, our regularized locally excited state $e^{-\ep H}\mathcal{O}(x)|0\lb$ can be written as a linear combination of infinitely many primary and descendant states. From this perspective, the parameter $\ep$ plays the role of an UV cut off for the selection of such states.

As pointed out in the remarkable paper \cite{kaplan}, the 4-pt function of two heavy operators $\mathcal{O}_B$ with two light operators $\mathcal{O}_A$ equals a thermal 2-pt function in the large $c$ limit as follows:
\be
\langle \mathcal{O}_B|\mathcal{O}_A(L)\mathcal{O}_A(0)|\mathcal{O}_B\rangle=\left(\frac{1}{\pi T_B}\sinh\left(\pi T_B L\right)\right)^{-2\Delta_A}
\label{2ptT}
\ee
with effective temperature determined in terms of the conformal dimension $\Delta_B$ of the heavy operator
\be
T_B=\frac{1}{2\pi}\sqrt{\frac{24}{c}\Delta_B-1}\,.
\ee
This observation follows from the old observation that conformal blocks exponentiate in the large $c$ limit \cite{BPZ}.

In fact, this result could be directly applied to the calculation of entanglement entropy in heavy excited states. This observable equals
\begin{equation}
  S_L=\lim_{n\to 1}\left[\frac{1}{1-n}\log\left(\langle \mathcal{O}_B|\sigma_n(L)\tilde{\sigma}_n(0)|\mathcal{O}_B\rangle\right)\right]\,. \label{ttt}
\end{equation}
Using the conformal dimension of the twist operators $\sigma_n$
\be
  \Delta_\sigma=\frac{c}{24}\left(n-\frac{1}{n}\right)\,,
\ee
and taking the limit $n\to 1$, we derive the large $c$ limit value for this entanglement entropy
\be
  S_L=\frac{c}{6}\log\left(\frac{\beta_B}{\pi}\sinh\left(\frac{\pi L}{\beta_B}\right)\right) \label{tttt}
\ee
where $\beta_B = T_B^{-1}$. This agrees with the entanglement entropy at finite temperature in two dimensional CFTs computed in \cite{CC} and also agrees with our expectation from the AdS/CFT correspondence. When finishing this work, we became aware that the authors of \cite{tom} got the same observation \eqref{ttt} and \eqref{tttt} independently. We refer readers to this paper for further analysis.

%%%%%%%%%%%%%%%%%%%%%%%%%%%%%%%%%%%%%%%%%
%%%%%%%%%%%%%%%%%%%%%%%%%%%%%%%%%%%%%%%%%
\section{Cross-ratios}\label{App:Details}
%%%%%%%%%%%%%%%%%%%%%%%%%%%%%%%%%%%%%%%%%
%%%%%%%%%%%%%%%%%%%%%%%%%%%%%%%%%%%%%%%%%

In this appendix we collect the explicit form of the exact cross-ratios used in the main text and determining the behaviour of the different quantum Renyi entropies and mutual information. For the map \eqref{MapSI}, they are given by
\bea
z_A&=&\frac{1}{2} \left(1+\frac{\cos \left(\frac{2 \pi  \epsilon }{\beta }\right)
   e^{\frac{2 \pi  (t-l)}{\beta }}-1}{\sqrt{\left(\cos
   \left(\frac{2 \pi  \epsilon }{\beta }\right) e^{\frac{2 \pi  (t-l)}{\beta
   }}-1\right)^2+\sin ^2\left(\frac{2 \pi
   \epsilon }{\beta }\right) e^{\frac{4 \pi  (t-l)}{\beta }}}}\right)\nn\\
\bar{z}_A&=&\frac{1}{2} \left(1+\frac{\cos \left(\frac{2 \pi  \epsilon }{\beta }\right)
   e^{-\frac{2 \pi  (t+l)}{\beta }}-1}{\sqrt{\left(\cos
   \left(\frac{2 \pi  \epsilon }{\beta }\right) e^{-\frac{2 \pi  (t+l)}{\beta
   }}-1\right)^2+\sin ^2\left(\frac{2 \pi
   \epsilon }{\beta }\right) e^{-\frac{4 \pi  (t+l)}{\beta }}}}\right)\label{CRSI}
\eea
For the evolution of the thermofield double with hamiltonian $H_R-H_L$ and maps \eqref{MapsLmR}, we have for the cylinder with a cut $B$
\bea
z_B&=&\frac{1}{2} \left(1-\frac{\cos \left(\frac{2 \pi
   \epsilon }{\beta }\right) e^{\frac{2 \pi  (t-l)}{\beta
   }}+e^{\frac{2 \pi  b}{\beta }}}{\sqrt{e^{\frac{4 \pi
   (t-l)}{\beta }}+2 \cos \left(\frac{2 \pi  \epsilon
   }{\beta }\right) e^{\frac{2 \pi  (t-l+b)}{\beta }}+e^{\frac{4 \pi  b}{\beta }}}}\right)\nn\\
\bar{z}_B&=&\frac{1}{2} \left(1-\frac{\cos \left(\frac{2 \pi
   \epsilon }{\beta }\right) e^{-\frac{2 \pi  (l+t)}{\beta
   }}+e^{\frac{2 \pi  b}{\beta }}}{\sqrt{e^{-\frac{4 \pi
   (l+t)}{\beta }}+2 \cos \left(\frac{2 \pi  \epsilon
   }{\beta }\right) e^{-\frac{2 \pi  (t+l-b)}{\beta }}+e^{\frac{4 \pi  b}{\beta }}}}\right)\label{CRBm}
\eea
and for the cut $A\cup B$ the ratios are
\begingroup
\everymath{\scriptstyle}
\scriptsize
\bea
z_{A\cup B}&=&\frac{1}{2} \left(1+\frac{e^{\frac{4 \pi
    (t-l)}{\beta }}-\left(1-e^{\frac{2
   \pi  b}{\beta }}\right) \cos \left(\frac{2 \pi  \epsilon }{\beta
   }\right) e^{\frac{2 \pi  (t-l)}{\beta }}-e^{\frac{2 \pi  b}{\beta }}}{\sqrt{\left(-2 \cos \left(\frac{2 \pi  \epsilon
   }{\beta }\right) e^{\frac{2 \pi  (t-l)}{\beta }}+e^{\frac{4 \pi
   (t-l)}{\beta }}+1\right) \left(e^{\frac{4 \pi  b}{\beta }}+2 \cos
   \left(\frac{2 \pi  \epsilon }{\beta }\right) e^{\frac{2 \pi
   (b-l+t)}{\beta }}+e^{\frac{4 \pi  (t-l)}{\beta }}\right)}}\right)\nn\\
\bar{z}_{A\cup B}&=&\frac{1}{2} \left(1+\frac{e^{-\frac{4
   \pi  (l+t)}{\beta }}-\left(1-e^{\frac{2
   \pi  b}{\beta }}\right) \cos \left(\frac{2 \pi  \epsilon }{\beta
   }\right) e^{-\frac{2 \pi  (l+t)}{\beta }}-e^{\frac{2 \pi  b}{\beta }}}{\sqrt{\left(e^{-\frac{4
   \pi  (l+t)}{\beta }}-2 \cos \left(\frac{2 \pi
   \epsilon }{\beta }\right) e^{-\frac{2 \pi  (l+t)}{\beta }}+1\right) \left(e^{\frac{4 \pi  b}{\beta }}+2
   \cos \left(\frac{2 \pi  \epsilon }{\beta }\right) e^{\frac{2 \pi
   (b-l-t)}{\beta }}+e^{-\frac{4 \pi  (l+t)}{\beta }}\right)}}\right)\label{CRABm}
\eea
\endgroup
When the evolution in the thermofield double is generated by the hamiltonian  $H_R+H_L$ and maps \eqref{MapsLpR}, the cross-ratios for the cylinder with cut $B$ are
\bea
z_B&=&\frac{1}{2} \left(1-\frac{\cos \left(\frac{2 \pi
   \epsilon }{\beta }\right) e^{\frac{2 \pi  (t-l)}{\beta
   }}+e^{\frac{2 \pi  (b+2t)}{\beta }}}{\sqrt{e^{\frac{4 \pi
   (t-l)}{\beta }}+2 \cos \left(\frac{2 \pi  \epsilon
   }{\beta }\right) e^{\frac{2 \pi  (3t-l+b)}{\beta }}+e^{\frac{4 \pi (b+2t)}{\beta }}}}\right)\nn\\
\bar{z}_B&=&\frac{1}{2} \left(1-\frac{\cos \left(\frac{2 \pi
   \epsilon }{\beta }\right) e^{-\frac{2 \pi  (l+t)}{\beta
   }}+e^{\frac{2 \pi  (b-2t)}{\beta }}}{\sqrt{e^{-\frac{4 \pi
   (l+t)}{\beta }}+2 \cos \left(\frac{2 \pi  \epsilon
   }{\beta }\right) e^{-\frac{2 \pi  (3t+l-b)}{\beta }}+e^{\frac{4 \pi  (b-2t)}{\beta }}}}\right)\label{CRBp}
\eea
whereas those for the cut $A\cup B$ are
\begingroup
\everymath{\scriptstyle}
\scriptsize
\bea
z_{A\cup B}&=&\frac{1}{2} \left(1+\frac{e^{\frac{4 \pi
    (t-l)}{\beta }}-\left(1-e^{\frac{2
   \pi  (b+2t)}{\beta }}\right) \cos \left(\frac{2 \pi  \epsilon }{\beta
   }\right) e^{\frac{2 \pi  (t-l)}{\beta }}-e^{\frac{2 \pi  (b+2t)}{\beta }}}{\sqrt{\left(-2 \cos \left(\frac{2 \pi  \epsilon
   }{\beta }\right) e^{\frac{2 \pi  (t-l)}{\beta }}+e^{\frac{4 \pi
   (t-l)}{\beta }}+1\right) \left(e^{\frac{4 \pi  (b+2t)}{\beta }}+2 \cos
   \left(\frac{2 \pi  \epsilon }{\beta }\right) e^{\frac{2 \pi
   (b-l+3t)}{\beta }}+e^{\frac{4 \pi  (t-l)}{\beta }}\right)}}\right)\nn\\
\bar{z}_{A\cup B}&=&\frac{1}{2} \left(1+\frac{e^{-\frac{4
   \pi  (l+t)}{\beta }}-\left(1-e^{\frac{2
   \pi  (b-2t)}{\beta }}\right) \cos \left(\frac{2 \pi  \epsilon }{\beta
   }\right) e^{-\frac{2 \pi  (l+t)}{\beta }}-e^{\frac{2 \pi  (b-2t)}{\beta }}}{\sqrt{\left(e^{-\frac{4
   \pi  (l+t)}{\beta }}-2 \cos \left(\frac{2 \pi
   \epsilon }{\beta }\right) e^{-\frac{2 \pi  (l+t)}{\beta }}+1\right) \left(e^{\frac{4 \pi  (b-2t)}{\beta }}+2
   \cos \left(\frac{2 \pi  \epsilon }{\beta }\right) e^{\frac{2 \pi
   (b-l-3t)}{\beta }}+e^{-\frac{4 \pi  (l+t)}{\beta }}\right)}}\right)\label{CRABp}
\eea
\endgroup

\end{appendix}
%\newpage
%%%%%%%%%%%%%%%%%%%%%%%%%%%%%%%%%%%%%%%
%%%%%%%%%%%%%%%%%%%%%%%%%%%%%%%%%%%%%%%

\end{document}